\documentclass[a4paper,fleqn,usenatbib]{mnras}

\usepackage{newtxtext,newtxmath}

\usepackage[T1]{fontenc}
\usepackage{ae,aecompl}
\usepackage{longtable}
\usepackage{hhline}

\usepackage{graphicx}	
\usepackage{amsmath}	
\usepackage{amssymb}	
\usepackage{lscape}
\usepackage{multirow}

\title[Lifetimes and f-values in Sc II]{Lifetime measurements and oscillator strengths in singly ionised scandium and the solar abundance of scandium}

\author[A. Pehlivan Rhodin et al.]{
A. Pehlivan Rhodin$^{1,2}$\thanks{E-mail:  asli.pehlivan@mah.se (APR)}$,$
M. T. Belmonte$^{3},$
L. Engstr\"om$^{4},$
H. Lundberg$^{4},$
H. Nilsson$^{1},$
\newauthor
H. Hartman$^{1,2},$
J. C. Pickering$^{3},$
C. Clear$^{3},$
P. Quinet$^{5,6}$,
V. Fivet$^{5}$,
and P. Palmeri$^{5}$\thanks{E-mail: patrick.palmeri@umons.ac.be (PP)}
\\
$^{1}$Lund Observatory, Lund University, PO Box 43, SE-221 00 Lund, Sweden\\
$^{2}$Materials Science and Applied Mathematics, Malm\"o University, 205 06 Malm\"o, Sweden \\
$^{3}$Physics Department, Blackett Laboratory, Imperial College London, London SW7 2BZ, UK\\
$^{4}$Department of Physics, Lund Institute of Technology, PO Box 118, SE-221 00 Lund, Sweden\\
$^{5}$Physique Atomique et Astrophysique, Universit\'e de Mons -- UMONS, 20 Place du Parc, B-7000 Mons, Belgium\\
$^{6}$IPNAS, Universit\'e de Li\`ege, B15 Sart Tilman, B-4000 Li\`ege, Belgium
}

\date{Accepted 2017 August 18. Received 2017 August 18; in original form 2017 July 12}

\pubyear{2017}

\begin{document}
\label{firstpage}
\pagerange{\pageref{firstpage}--\pageref{lastpage}}
\maketitle

\begin{abstract}
The lifetimes of 17 even-parity levels (3d5s, 3d4d, 3d6s, and 4p$^2$) in the region 57743-77837 cm$^{-1}$ of singly ionised scandium (\ion{Sc}{ii}) were measured by two-step time-resolved laser induced fluorescence spectroscopy. Oscillator strengths of 57 lines from these highly excited upper levels were derived using a hollow cathode discharge lamp and a Fourier transform spectrometer. In addition, Hartree--Fock calculations where both the main relativistic and core-polarisation effects were taken into account were carried out for both low- and high-excitation levels. There is a good agreement for most of the lines between our calculated branching fractions and the measurements of \citet{law89} in the region 9000-45000 cm$^{-1}$ for low excitation levels and with our measurements for high excitation levels in the region 23500-63100 cm$^{-1}$. This, in turn, allowed us to combine the calculated branching fractions with the available experimental lifetimes to determine semi-empirical oscillator strengths for a set of 380 E1 transitions in \ion{Sc}{ii}. These oscillator strengths include the weak lines that were used previously to derive the solar abundance of scandium. The solar abundance of scandium is now estimated to ${\rm log}~\epsilon_{\sun}=3.04\pm0.13$ using these semi-empirical oscillator strengths to shift the values determined by \citet{sco15}. The new estimated abundance value is in agreement with the meteoritic value (${\rm log}~\epsilon_{\text{met}}=3.05\pm0.02$) of \citet{lod09}.

\end{abstract}




\section{Introduction}

The iron-group elements ($21 \leq Z \leq 28$) are produced during supernova type Ia explosions, while supernova type II explosions are responsible for the formation of $\alpha$-elements such as Mg, Si, S. The even-$Z$ nuclei such as S, Ca, Ti, Cr, and Fe have higher cosmic abundance compared to the odd-$Z$ nuclei located in between because of the consecutive capture of $\alpha$-particles. The production of odd-$Z$ elements is not as well understood, and does not follow the abundance trends of the $\alpha$-elements, indicating non-common production mechanisms. In recent years, this has caused an increasing interest in the odd-$Z$ iron-peak elements in astrophysics. Abundance determinations in stars constrain the stellar evolution and supernova explosion models \citep{pag09}.  Moreover, transitions from highly excited levels have additional diagnostic value, since they can be used to benchmark non local thermodynamical equilibrium (NLTE) modelling of stellar atmospheres. Besides the development of 3D hydrodynamic model atmospheres, a trustworthy NLTE treatment is the current challenge for accurate stellar abundances. High-precision atomic data for selected lines are important for this development \citep{lin12}. \\
\indent In the case of scandium ($Z=21$), a realistic 3D NLTE solar atmosphere model has been used by \citet{sco15} to revise the solar abundance of scandium resulting in a photospheric value in significant disagreement with the meteoritic abundance \citep{lod09}. \citet{sco15} used experimental transition probabilities of five \ion{Sc}{i} and nine \ion{Sc}{ii} lines determined by \citet{law89}. The latter authors combined their measured branching fractions with the time-resolved laser induced fluorescence (TR-LIF) lifetimes of \citet{mar88} to obtain absolute $A$-values for transitions depopulating 51 levels in \ion{Sc}{i} and 18 levels in \ion{Sc}{ii}. In \citet{mar88}, only three highly-excited even-parity levels of \ion{Sc}{ii}, belonging to ${\rm 3d4d~^3G}$, were measured. Older lifetime measurements in singly ionised scandium have focussed on lower excited  odd-parity ${\rm 3d4p}$ and ${\rm 4s4p}$ levels \citep{buc71,arn76,pal76,vog85}. On the theoretical side, the most recent calculations of E1 oscillator strengths in \ion{Sc}{ii} are \cite{ruc14} and \citet{kur15}. \\
\indent The main goal of the present work is to provide a new set of experimental $f$-values for transitions depopulating the highly-excited even-parity levels in \ion{Sc}{ii}, and new calculations for both low- and high-excitation levels and lines. Descriptions of our measurements are presented in Section~2 and 3. The theoretical method used for the calculation of the radiative parameters is described in Section~4. In Section~5, our results are presented and compared to data available in the literature. The consequence of the proposed set of oscillator strengths on the solar abundance of scandium is discussed in Section~6. Finally, our conclusions are given in Section~7. 

\section{Lifetime measurements}

\begin{figure*}
\centering
	\includegraphics[width=\textwidth]{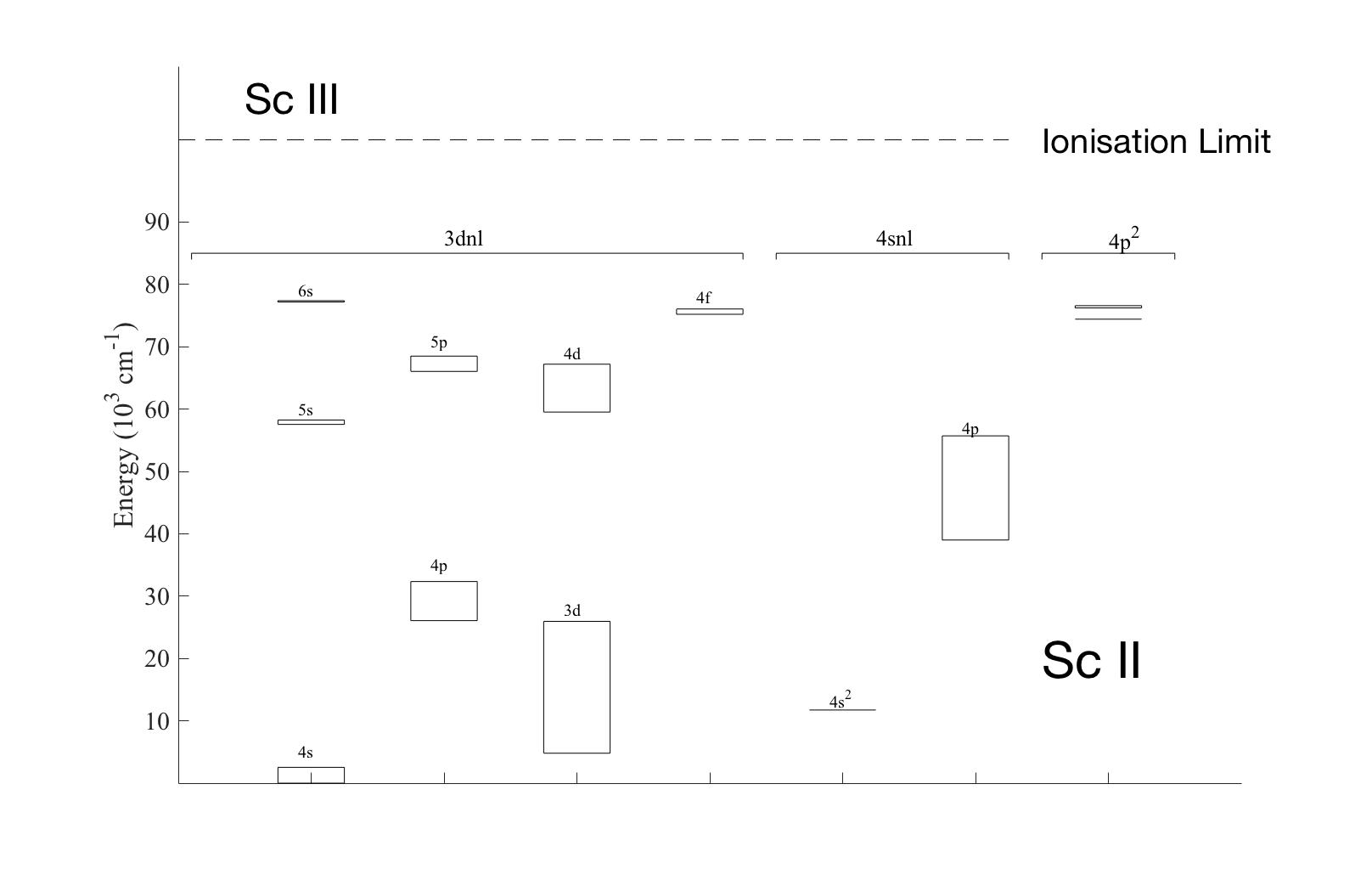}
    \caption{Partial energy level diagram of \ion{Sc}{ii}, the energy level values are from \citet{joh80}. Each box consists of several levels.}
    \label{energy_level}
\end{figure*}

The experimental set-up for the two-step Time-Resolved Laser Induced Fluorescence  (TR-LIF) measurements at the Lund High Power Laser Facility has been described in detail by \citet{en14} and \citet{lun16}. For an overview we refer to Figure~1 in \citet{lun16}, and here we give only the most important details. A frequency doubled Nd:YAG laser (Continuum Surelite) with 10 ns pulses was used to produce the free scandium ions by focusing the light on a rotating solid scandium sample in a vacuum chamber with a pressure around $10^{-4}$ mbar. The ions in the plasma cone were crossed by two laser beams, a few mm above the solid sample, generating the two-step excitations. The fluorescence signal was detected in a direction perpendicular to both the ablation and excitation lasers.\\
\indent For the first step (4s-4p), we used a Continuum Nd - 60 dye laser with either DCM or Pyridine 2 dyes. The 10 ns long pulses were frequency doubled using a KDP crystal, giving the wavelengths needed for the first step. The second laser system excited the final high energy levels. It consists of a frequency doubled Continuum NY-82 Nd:YAG laser pumping a Continuum Nd - 60 dye laser with either DCM or Oxacin dye for wavelengths below or above 660~nm, respectively. The pulse length was reduced from 10 ns to less than one ns by stimulated Brillouin scattering. The output was frequency doubled using a KDP crystal and, where higher energy was needed, tripled with a BBO crystal.  \\
\indent For two step excitation, the timing between the pulses is crucial. For this purpose, a delay generator ensures that the second step is timed to when the population of the intermediate state is at its flat maximum as determined by observing the decay of this level in another channel, see Figure 2 in \citet{lun16}.  \\
\indent The fluorescence emitted by the scandium ions was filtered by a 1/8 m grating monochromator with its 0.28 mm wide entrance slit oriented parallel to the excitation laser beams. This fluorescence light was recorded using a fast micro-channel-plate photomultiplier tube (Hamamatsu R3809U) and digitised using a Tektronix DPO 7254 oscilloscope with 2.5 GHz analog bandwidth. We used the second spectral order with a 0.5 nm observed line width for all measurements. The excitation laser pulse shape was recorded simultaneously using a fast photo diode and digitised by another channel of the oscilloscope. All decay curves were averaged over 1000 laser pulses and analysed using the DECFIT software \citep{pal08} by fitting a single exponential function convoluted by the measured shape of the second-step laser pulse and a background function to the observed decay.\\
\indent The excitation schemes of the measured \ion{Sc}{ii} levels are presented in Table 1. This table shows the intermediate levels and their excitation wavelengths, the final levels and their excitation wavelengths from the intermediate levels together with the detection channel level and wavelength. For the levels 4d ${}^3$S$_1$, 4d ${}^1$D$_2$ and 4p$^2$ ${}^3$P$_2$, it was possible to record the the decay in more than one channel. We did not find any differences in the lifetimes obtained from the different channels. \ion{Sc}{ii} is a complex spectrum with a dense level structure, as shown in Figure~\ref{energy_level}. Line blending can be caused by cascades or fluorescence from the intermediate level as discussed by \citet{lun16}. For all measurements, we investigated if there was a line blend affecting the recorded curves.  Due to the small spectral width of the laser compared to the energy level separations, we avoid exciting multiple levels. \\
\indent To investigate any possible saturation effects in the second step excitation, a set of neutral density filters was placed in the excitation beam. The delay between the ablation and first excitation pulse, the geometrical alignment of the lasers in respect to the target as well as the intensity of the ablation laser were varied to test time-of-flight effects. No systematic effects were observed.\\
\indent As discussed in \citet{pal08}, the weighting of individual data points, hence the purely statistical uncertainty in the fitted lifetime, is difficult to estimate accurately because the digitising process is not strictly a counting measurement. However, extensive tests have shown that even for weak lines the dominating factor is the variation between different measurements. The uncertainty in Table~\ref{life} represents the uncertainty of 10-20 measurements performed over several days. The difference between subsequent curves is significantly lower than the quoted uncertainty, usually less than 1$\%$.

\section{Branching fraction measurements}

A water-cooled hollow cathode discharge lamp (HCL) was used to produce the free scandium ions. The lamp has an iron cathode with anodes on each side, separated by glass cylinders. A small piece of scandium was placed in the cathode. We used argon, with a pressure of 0.3 Torr, as a buffer gas and applied currents ranging from 0.2 to 0.5 A. These measurements at different currents are very important to find and compensate for self-absorption effects. If self-absorption is not treated correctly, the measured relative line intensity may be less than the true intensity of the line. This in turn changes the branching fraction which is essential to derive oscillator strengths. Self-absorption was observed in the case of the ${\rm 3d4d~^3D_3}$, ${\rm 3d4d~^3S_1}$, and ${\rm 3d4d~^3P_2}$ levels, and the affected lines were corrected. More details on this procedure can be found in \cite{peh15}. \\
\indent The spectra were recorded with the vacuum ultraviolet Fourier transform spectrometer (VUV FTS) at the Blackett Laboratory, Imperial College London \citep{pick02} in the interval $23500 - 63100$ $\text{cm}^{-1}$ ($425-158$ nm) using a resolution of 0.039 cm$^{-1}$. We used two different photomultiplier tube detectors: Hamamatsu R7154 and  R11568, the latter with a UG5 filter. Each scandium measurement consists of 12 co-added scans. To determine the relative response functions of the system, we used standard lamps: a tungsten filament lamp ($800-300$ nm) and a deuterium lamp ($410-116$ nm) for the wavelength region ($425-210$ nm), and a deuterium standard lamp alone for the region ($317-158$ nm). The tungsten lamp was calibrated by the UK National Physical Laboratory and the deuterium lamp by Physikalisch-Technische Bundesanstalt, in  Berlin. In the region where the lamps overlap, the response functions were placed on the same relative scale. We recorded the spectrum of the calibration lamps immediately before and after each scandium measurement. The HCL and the calibration lamps were placed at the same distances from the FTS, and a mirror was used to select the light source without moving the lamps. \\
\indent In astrophysics, oscillator strengths ($f$-values) or $\log(gf)$ values are the parameters used for abundance analysis. The $f$-value is proportional to the transition probability for E1 transitions by
\begin{equation}
f=\frac{g_{\text{u}}}{g_{\text{l}}} \lambda^2 A_{\text{ul}}1.499 \times 10^{-16},
\end{equation}
where $g_{\text{u}}$ is the statistical weight of the upper level, $g_{\text{l}}$ the statistical weight of the lower level, $\lambda$ the wavelength of the transition in \AA, and $A_\text{ul}$ the transition probability in s$^{-1}$. \\
\indent The transition probability is related to the branching fraction ($BF$) and the lifetime of the upper level ($\tau_{\text{u}}$). It can be derived using
\begin{equation}
A_{\text{ul}} =\frac{BF_\text{ul}} {\tau_{\text{u}}}.
\end{equation}
We obtained the lifetimes of the upper levels from our measurements, as discussed in Section 2. The $BF$ is the parameter we measure and it is defined as the transition probability of the line, $A_{\text{ul}}$, divided by the sum of transition probabilities of all lines from the same upper level;
\begin{equation}
BF_{\text{ul}} = \frac{A_\text{ul}}{\sum_i A_{\text{ui}}} = \frac{I_\text{ul}}{\sum_i I_{\text{ui}}}. 
\end{equation}
Since all lines emanate from the same upper level, the transition probability is proportional to the line intensity, $I_{\text {ul}}$, which for FTS spectra is proportional to photon flux \citep{book1}. Therefore, we derived $BF$s from calibrated intensity ratios in our measurements.  All lines were identified using the analysis of \citet{joh80}. The intensities of the observed lines were determined by fitting Gaussian line profiles using GFit \citep{lars98, lars14}. \\
\indent The uncertainty of the $A$-value, and thus of the $f$-value, arises from the uncertainty in the upper level lifetime and the uncertainty of the $BF$. The latter includes the uncertainty in the intensity calibration procedure and the uncertainty in the measured line intensity, including the self-absorption correction. The uncertainties of the integrated line intensities were determined using GFit. The relative uncertainties are as low as 0.1\% for strong lines and 4\% on average. However, for two weak lines the uncertainty is as large as $20\%$. The uncertainty in the calibration using the tungsten lamp is 2.2\% and the uncertainty using the deuterium lamp is 8.6\% for the region $425-210$ nm and 9.9\% between $317$ and $158$ nm. These calibration lamp uncertainties include the calibration uncertainty and the variation resulting from the repeated measurements made before and after all scandium scans. The uncertainties of the radiative lifetimes are given in Table \ref{life}. Finally, we were not able to observe the weakest lines from the investigated level. However, we included their contributions as residuals with derived theoretical $BF$s from our calculations. The residual $BF$s are less than $7\%$ for all levels. The uncertainties in the residuals were estimated to $50\%$ and included in the error budget. The final uncertainties in the oscillator strengths are presented in Table \ref{exp} and were derived from the individual contributions using the method described by \citet{siks02}.

\section{Radiative parameter calculations}

To calculate branching fractions and the oscillator strengths in \ion{Sc}{ii}, we used the relativistic Hartree--Fock (HFR) method implemented in the Cowan's suite of atomic structure computer codes \citep{cow81}. It is modified by including a pseudo-potential and a correction to the electric dipole operator that take into account the core-polarisation effects  giving rise to the HFR+CPOL technique \citep{qui99}. \\
\indent In this study, the valence-valence correlation was included using the following configuration interaction (CI) expansions: 
${\rm 3d 4s}$ + ${\rm 3d 5s}$ + ${\rm 3d 6s}$ + ${\rm 3d 7s}$ +
${\rm 3d^2}$ + ${\rm 3d 4d}$ + ${\rm 3d 5d}$ + ${\rm 3d 6d}$ +
${\rm 3d 7d}$ + ${\rm 3d 5g}$ + ${\rm 3d 6g}$ + ${\rm 3d 7g}$ +
${\rm 4s^2}$+${\rm 4s 5s}$ + ${\rm 4s 6s}$ + ${\rm 4s 7s}$ + ${\rm 4s 4d}$ +
${\rm 4s 5d}$ + ${\rm 4s 6d}$ + ${\rm 4s 7d}$ + ${\rm 4s 5g}$ +
${\rm 4s 6g}$ + ${\rm 4s 7g}$ + ${\rm 4p^2}$ + ${\rm 4d^2}$ +
${\rm 4f^2}$ + ${\rm 4p 4f}$ for the even parity; ${\rm 3d 4p}$ +
${\rm 3d 5p}$ + ${\rm 3d 6p}$ + ${\rm 3d 7p}$ + ${\rm 3d 4f}$ +
${\rm 3d 5f}$ + ${\rm 3d 6f}$ + ${\rm 3d 7f}$ + ${\rm 3d 6h}$ +
${\rm 3d 7h}$ + ${\rm 4s 4p}$ + ${\rm 4s 5p}$ + ${\rm 4s 6p}$ +
${\rm 4s 7p}$ + ${\rm 4s 4f}$ + ${\rm 4s 5f}$ + ${\rm 4s 6f}$ +
${\rm 4s 7f}$ + ${\rm 4s 6h}$ + ${\rm 4s 7h}$ + 
${\rm 4p 4d}$ + ${\rm 4d 4f}$ for the odd parity. \\
\indent Regarding the core-polarisation effects, a \ion{Sc}{iv} ${\rm 3p^6}$ closed-subshell ionic core was considered where the dipole polarisability, $\alpha_\text{d}= 2.129$ $a_0^3$  was taken from the RRPA calculations of \citet{joh83} and a cut-off radius of 1.17 $a_0$ was estimated as the HFR mean radius of the outermost 3p orbital, $\left\langle3p|r|3p\right\rangle_{\text{HFR}}$.\\
\indent During a least-squares-fit procedure, we adjusted some radial integrals to minimise the discrepancies between the hamiltonian eigenvalues and  the experimental energy levels taken from the NIST Atomic Spectra Database \citep{kra15}. The latter are based on the term analysis originally carried out by \citet{rus27} and later revised by \citet{neu70} and by \citet{joh80}. There are 168 levels belonging to the configurations
${\rm 3d 4s}$, ${\rm 3d^2}$, ${\rm 3d 4p}$, ${\rm 4s 4p}$, 
${\rm 3d 5s}$, ${\rm 3d 4d}$, ${\rm 3d 5p}$, ${\rm 4p^2}$,
${\rm 3d 4f}$, ${\rm 3d 6s}$, ${\rm 4s 5s}$, ${\rm 3d 5d}$,
${\rm 4s 4d}$, ${\rm 3d 5f}$, ${\rm 3d 5g}$, ${\rm 3d 7s}$,
${\rm 3d 6d}$, and ${\rm 3d 6f}$. The average energies, $E_{\text{av}}$, of the above-mentioned known configurations along with their direct, $F^k$, exchange, $G^k$, electrostatic and spin-orbit, $\zeta$, radial parameters were considered in the fit of the energy levels. The {\it ab initio} and fitted parameter values are reported in Tables~\ref{epar} and \ref{opar} for the even and
odd configurations, respectively. The spin-orbit integrals not presented in these tables were fixed to their HFR+CPOL values. The other Slater integrals, including the CI $R^k$ parameters, not reported here, were fixed to 80\% of their {\it ab initio} values to account for missing CI effects \citep{cow81}. The average deviations of the least-squares-fits were 157 cm$^{-1}$ for the 93 even-parity experimental levels and 65 cm$^{-1}$ for the 75 odd-parity experimental levels.

\section{Results and discussion}

\begin{figure}
	\includegraphics[width=\columnwidth]{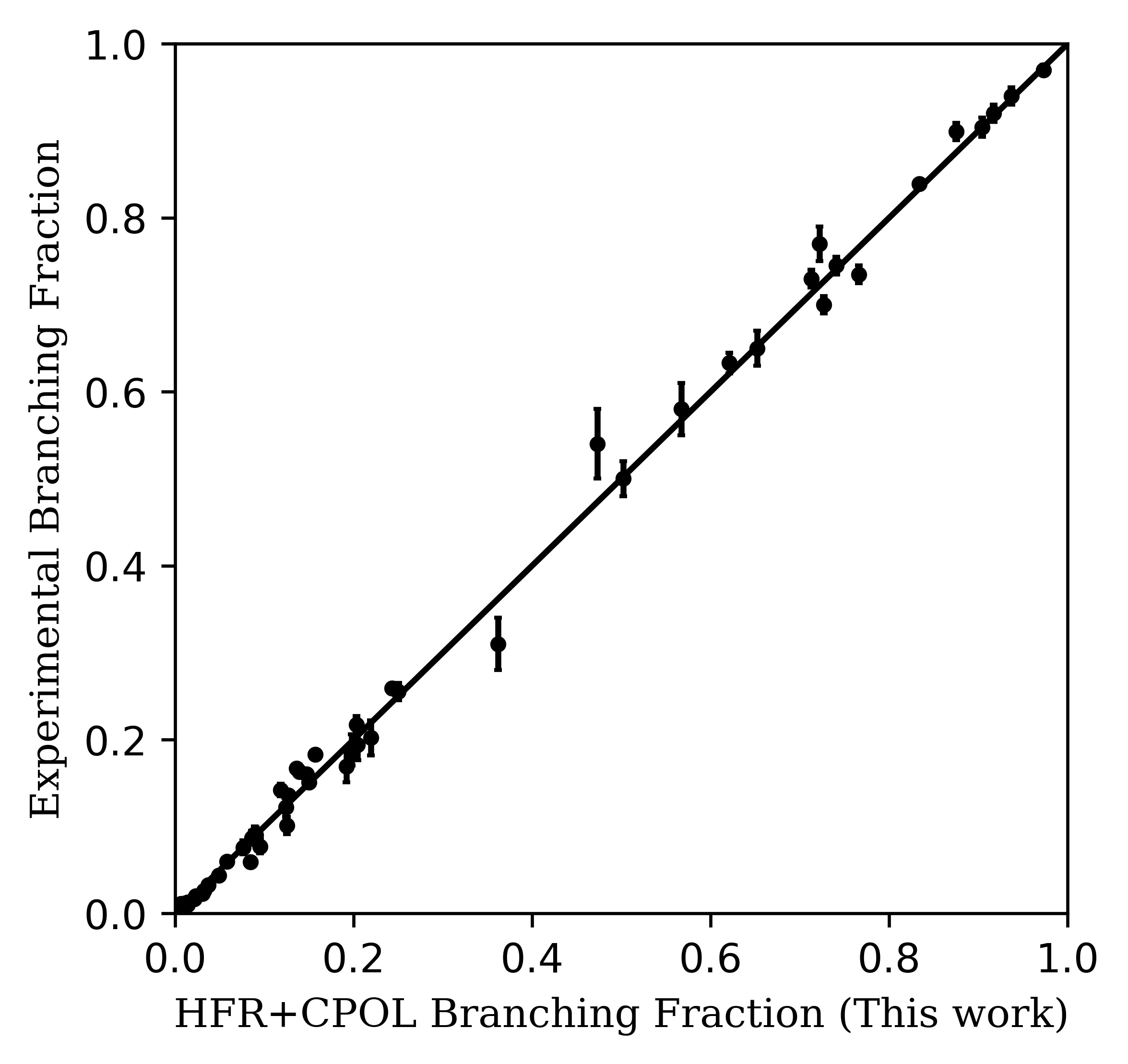}
    \caption{Comparison between the HFR+CPOL branching fractions of this work
    and the experimental values of \citet{law89}. The straight line of
    equality has been drawn.}
    \label{comp1}
\end{figure}

\begin{figure}
	\includegraphics[width=\columnwidth]{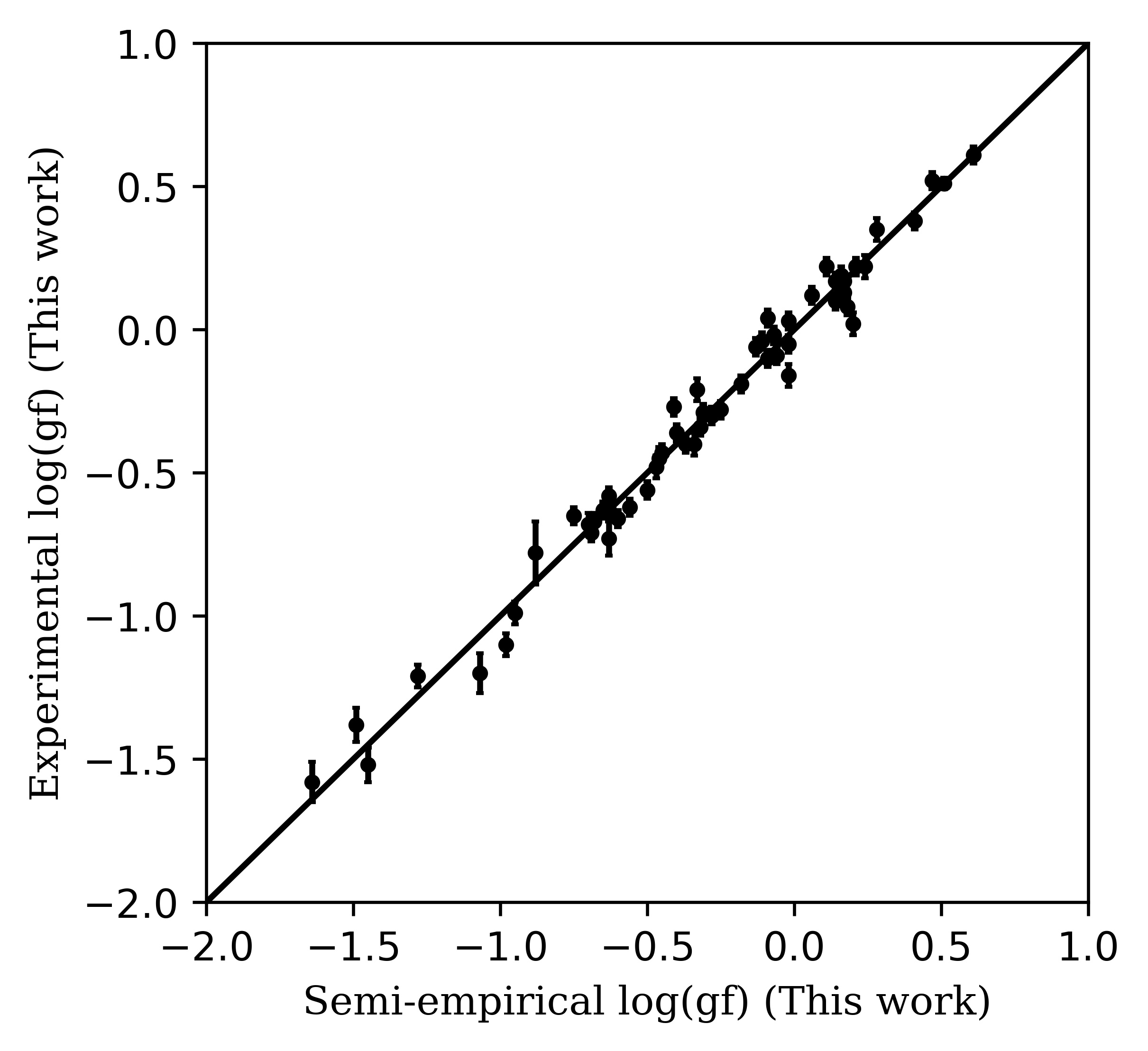}
   \caption{Comparison between the oscillator strengths determined by
    the combination of the HFR+CPOL branching fractions and the TR-LIF
    lifetimes of this work and the experimental oscillator strengths derived in this work. The straight line of equality has been drawn.}
    \label{comp6}
\end{figure}

\begin{figure}
\includegraphics[width=\columnwidth]{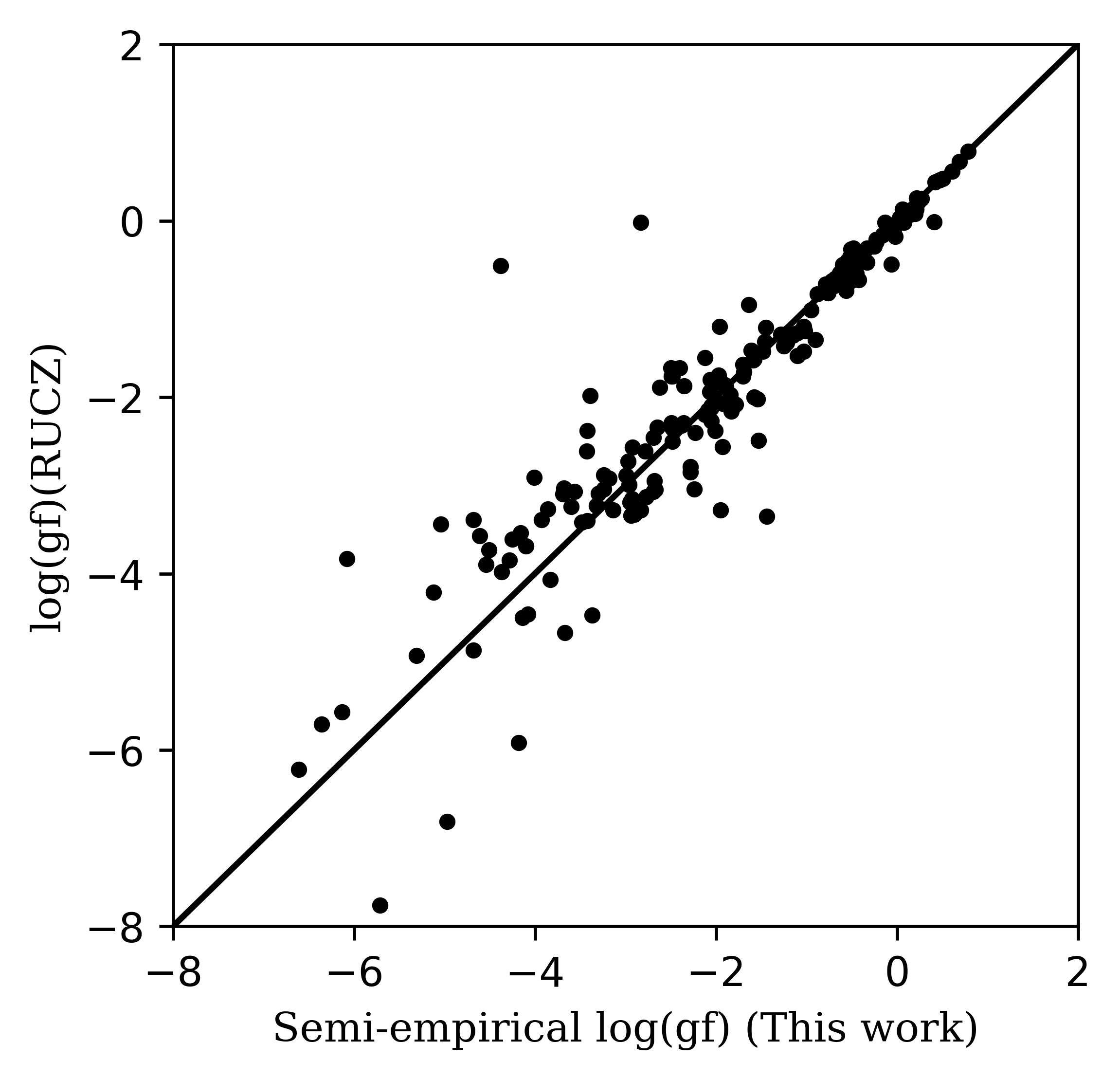}
    \caption{Comparison between the oscillator strengths determined by
    the combination of the HFR+CPOL branching fractions and the TR-LIF
    lifetimes (This Work) and the semi-empirical oscillator strengths calculated by \citet{ruc14}
    (RUCZ). The straight line of equality has been drawn.}
    \label{comp3}
\end{figure}

\begin{figure}
	\includegraphics[width=\columnwidth]{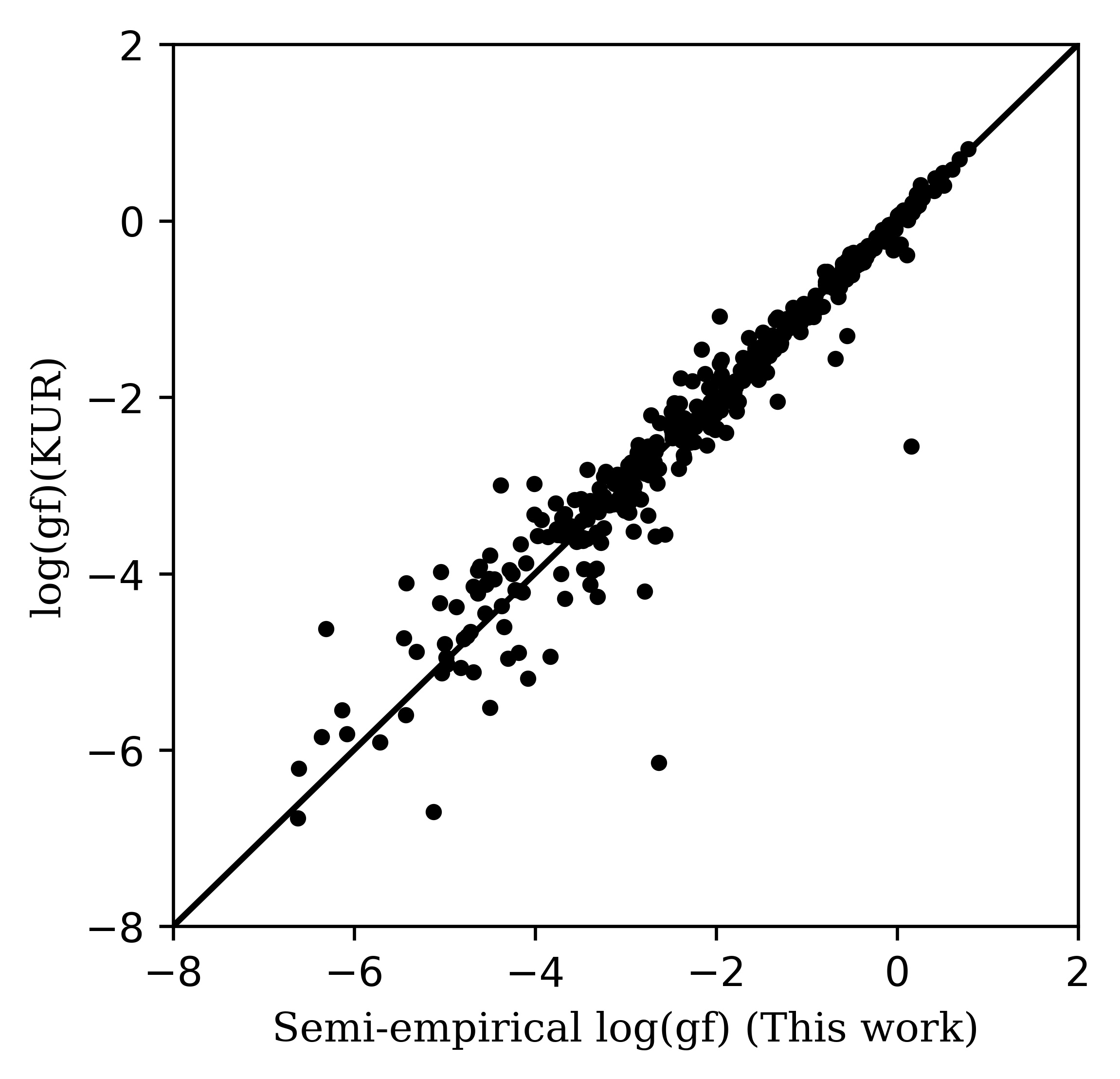}
    \caption{Comparison between the oscillator strengths determined by
    the combination of the HFR+CPOL branching fractions and the TR-LIF
    lifetimes (This Work) and the oscillator strengths calculated by \citet{kur15}
    (KUR). The straight line of equality has been drawn.}
    \label{comp4}
\end{figure}

Table~\ref{life} compares our TR-LIF and HFR+CPOL lifetimes with other experimental values from the literature \citep{buc71,arn76,pal76,vog85,mar88}, the Hartree-Fock values calculated by \citet{kur15} and the lifetimes deduced from the semi-empirical oscillator strengths calculated by \citet{ruc14}. On average, our HFR+CPOL lifetimes are shorter than the measurements for the odd-parity levels and longer for the even-parity levels. The discrepancies range from a few percent to about 20\%, except for the even-parity levels ${\rm 4p^2~^1D_2}$ and ${\rm 3d6s~^3D_3}$ 
where they reach 57\% and 49\%, respectively. In the former case, this state is strongly mixed (our calculation gives 36\% ${\rm 4p^2~^1D_2}$ + 36\% ${\rm 4s4d~^1D_2}$ + 23\% ${\rm 3d6s~^1D_2}$) and an important decay channel (${\rm 4p^2~^1D_2 \rightarrow 3d4p~^1D^o_2}~~BF=0.0713$) is affected by cancellation (the cancellation factor as defined by \citet{cow81} is less than 5\%) that could explain the over estimated lifetime. Concerning ${\rm 3d6s~^3D_3}$ level, no such argument could explain the observed disagreement. The beam-foil measurements of \citet{buc71} can be rejected for the levels ${\rm 3d4p~^3D^o_3,^1P^o_1,^1F^o_3}$ as previously stated by \citet{mar88} due to blending problems. \\
\indent The calculations by \citet{kur15} show roughly the same systematic discrepancy with experiment (lifetimes shorter for the odd parity and longer for the even parity) as our HFR+CPOL calculations. Although the calculation of \citet{kur15} shows a better agreement than HFR+CPOL for certain 3d4d levels (${\rm ^3F_{2,4}}$, ${\rm ^1D_2}$, ${\rm ^3P_2}$), it does not solve the theory-experiment disagreements observed for the levels ${\rm 4p^2~^1D_2}$ and ${\rm 3d6s~^3D_3}$. The parametric calculation of \citet{ruc14} agrees with our HFR+CPOL model within 10\% including all levels. Unfortunately, no lifetime value can be deduced from \citet{ruc14} for the levels ${\rm 4p^2~^1D_2}$ and ${\rm 3d6s~^3D_3}$. Concerning the level ${\rm 3d4d~^3G_3}$, our TR-LIF measurement is slightly lower than the one of \citet{mar88} although the error bars do overlap. \\
\indent For all 3d4p levels, our HFR+CPOL model and the parametric calculation of \citet{ruc14} are closer to the measurement of \citet{mar88}. The excellent agreement between \citet{mar88} and \citet{ruc14} is not surprising as the latter adjusted the dipole transition integrals to the oscillator strengths determined from the branching fraction measurements of \citet{law89} combined with the lifetime measurements of \citet{mar88}. For most of the higher levels, the lifetimes calculated by \citet{kur15} are closer to our measurements than those of  \citet{ruc14}.\\
\indent Although there is a systematic discrepancy between the theoretical and experimental lifetimes, we find a better agreement when comparing our calculated $BF$s with the experimental values. For the high excitation lines, measured in this work, the averaged $BF$ ratio is $1.02\pm0.16$ with respect to the calculated values. Similarly, Figure~\ref{comp1} shows the good agreement between $BF$s computed in this study using the HFR+CPOL method and the measurements by \citet{law89}. Here, the averaged $BF$ ratio is $0.98\pm0.20$. Based on these comparisons, the calculated $BF$s were combined with our TR-LIF lifetimes and those of \citet{mar88} to determine rescaled transition probabilities and oscillator strengths. \\
\indent In Table~\ref{exp}, we present our experimental $\log(gf)$ values, together with the measured $BF$s, the uncertainties and the corresponding rescaled theoretical oscillator strengths, $\log(gf)_{\text{resc}}$. Figure~\ref{comp6} illustrates the final agreement between our experimental $\log(gf)$ values and the calculated $\log(gf)_{\text{resc}}$. Table~\ref{gfsc2} summarises our calculated radiative parameters along with the weighted transition probabilities ($gA$), the weighted oscillator strengths in the log scale ($\log(gf)$), the HFR+CPOL branching fractions ($BF$), and the cancellation factor ($CF$) as defined by \citet{cow81}. \\
\indent Our rescaled theoretical oscillator strengths are compared to the semi-empirical values calculated by \citet{ruc14} in Figure~\ref{comp3}. As expected, the scatter increases for the weak lines, i.e. the transitions with $\log(gf) \lesssim -1$, where cancellation effects could be an issue. For instance, the transition ${\rm 3d4p~^3P^o_2 - 4p^2~^3P_2}$ labelled in Table~\ref{gfsc2} $76589(e)2 - 29824(o)2$ has a very low cancellation factor ($CF=0.001$) that indicates a strong cancellation effect in our HFR+CPOL line strength calculation. Indeed, the rescaled oscillator strength for that transition is $\log(gf)_{\text{resc}}=-2.83$ which is three orders of magnitude lower (in the linear scale) than the value predicted by \citet{ruc14} ($\log(gf)=-0.02$). On the other hand, a transition for which the cancellation effects in our model is not an issue ($CF>0.05$) such as ${\rm 3d4p~^3F^o_3 - 3d4d~^3F_4}$ ($63529(e)4 - 27602(o)3$) has an oscillator strength predicted by \citet{ruc14} ($\log(gf)=-3.35$) that is two orders of magnitude lower than our rescaled value ($\log(gf)_{\text{resc}}=-1.44$). This could indicate a strong cancellation effect in their calculation. Unfortunately, they did not estimate any cancellation factors. For the strongest transitions, i.e. ${\log(gf)}\gtrsim -1$, the mean scatter drops to about 20\% in the linear scale. \\
\indent In Figure~\ref{comp4}, our semi-empirical values are compared to the calculation of \citet{kur15} where a similar global correlation is observed. The mean scatter in this case is also found to be $\sim$20\% for transitions with $\log(gf) \gtrsim -1$ and increases for weaker lines. Here again, the cancellation factors are not available in Kurucz's database \citep{kur15}. But, for example, our predicted strong line ${\rm 3d5p~^3F^o_3 - 3d6s~^3D_3}$ ($77387(e)3-66564(o)3$) with $\log(gf)_{\text{resc}}=0.16$ and $CF=0.379$ is certainly affected by a strong cancellation effect in the calculation of \citet{kur15} dramatically lowering its oscillator strength to $\log(gf)=-2.56$. \\
\indent Based on the differences between different sets of $BF$s discussed above and including the uncertainties of the experimental lifetimes, we estimate the accuracy of the rescaled theoretical $f$-values to be $10\%$ for the strong lines and $15-20\%$ for other lines.

\section{Consequence on the Solar abundance of scandium}

\begin{figure}
	\includegraphics[width=\columnwidth]{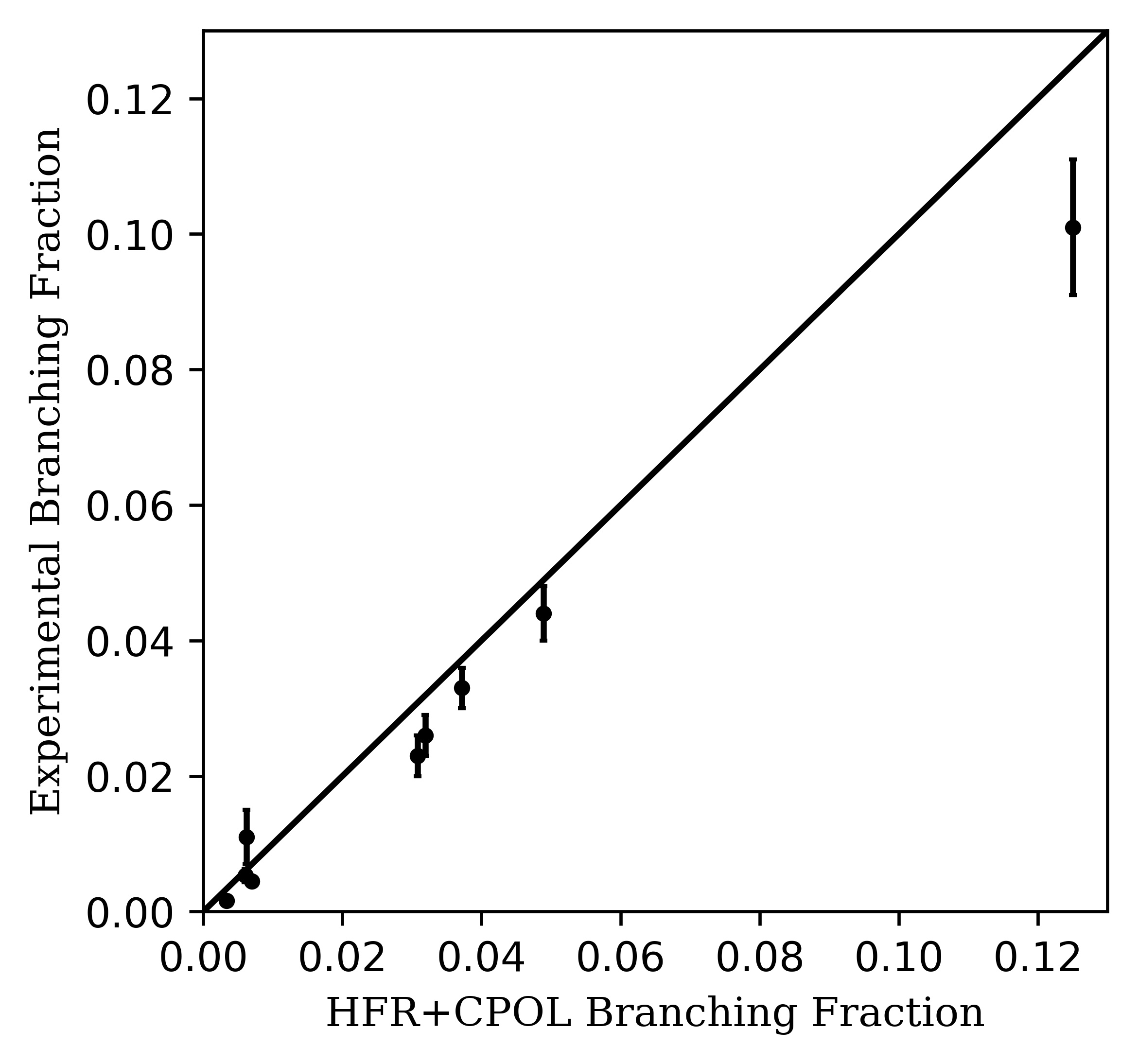}
    \caption{Comparison between the HFR+CPOL branching fractions 
    and the experimental values of \citet{law89} for the \ion{Sc}{ii} lines used in the determination of the solar scandium abundance. The straight line of
    equality has been drawn.}
    \label{comp2}
\end{figure}

\citet{sco15} have redetermined the solar abundances of the iron-peak elements employing a 3D model atmosphere that takes into account departures from the local thermodynamic equilibrium.
However, the significant discrepancy between the photospheric and the meteoritic abundances \citep{lod09} still remains for scandium with ${\rm log}~\epsilon_{\sun}=3.16\pm0.04$ and ${\rm log}~\epsilon_{\text{met}}=3.05\pm0.02$. \\
\indent The \ion{Sc}{ii} lines used in \citet{sco15} for the determination of the solar abundance of scandium are presented in Table~\ref{abun}. These lines are from low-excited levels measured by \citet{law89} but not included in our present study. The third column of this table contains the oscillator strengths deduced from the $A$-values of \citet{law89} used by \citet{sco15} to determine the photospheric abundances listed in sixth column. They are compared to our rescaled oscillator strengths reported in the fourth column and the differences between the two values are given in the log scale in the fifth column. For this set of solar lines, our oscillator strengths are systematically larger than those of \citet{law89} by $\sim$0.1 dex on average, if we exclude the transition ${\rm 3d^2~^1D_2 - 3d4p~^1D^o_2}$ for which our $f$-value is affected by strong cancellation effects. Column seven in Table~\ref{abun} gives the  abundances obtained from each line, assuming they are all lying on the linear part of the curve of growth (see the upper left panel of Figure 3 in \citet{sco15}), with our new $gf$-values. \\
\indent The weighted average along with the corresponding weighted standard deviation of the abundance were determined using the weights of \citet{sco15}, reported in the last column of Table~\ref{abun}. Their weights range from one to three and are based on the line quality for abundance determination. Discarding the line ${\rm 3d^2~^1D_2 - 3d4p~^1D^o_2}$ from the mean estimate, one obtains ${\rm log}~\epsilon_{\text{cor}}=3.04\pm0.13$  (where the second number is the standard deviation) for the corrected photospheric abundance, now in good agreement with the meteoritic value of \citet{lod09}. Even if we reject the transition ${\rm 3d^2~^3F_4 - 3d4p~^3F^o_3}$ for which there is a factor of two difference between our rescaled $f$-value and the experimental
value of \citet{law89}, the mean ${\rm log}~\epsilon_{\text{cor}}=3.10\pm0.05$ is still in accord with the meteoritic value. Moreover, considering the full line set does not change the agreement (${\rm log}~\epsilon_{\text{cor}}=3.07\pm0.17$). Finally we note that, all these weighted average abundances agree within the mutual error bars with the value determined by \citet{sco15} using only \ion{Sc}{i} lines (${\rm log}~\epsilon=3.14\pm0.09$).\\
\indent Replacing our $f$-value set by the one of \citet{kur15} will not change this accord either (${\rm log}~\epsilon_{\text{kur}}=3.10\pm0.09$). This is not the case for the set of \citet{ruc14}. Indeed, the photospheric abundance would be estimated significantly too high with respect to the meteoritic value, i.e. ${\rm log}~\epsilon_{\text{ruc}}=3.44\pm0.31$. Even if the transition  ${\rm 3d^2~^3F_4 - 3d4p~^3F^o_3}$ for which the oscillator strength calculated by \citet{ruc14} ($\log(gf)_{\text{ruc}}=-3.28$) is one order of magnitude lower than the experimental value of \citet{law89} is excluded, this would not significantly improve the situation (${\rm log}~\epsilon_{\text{ruc}}=3.29\pm0.01$).  \\
\indent It should be noted, however, that the lines used for these studies are weak, see Figure~\ref{comp2}. Their $BFs$ are less than $5\%$ except for $\lambda 566.904$ having $\sim10\%$. These small $BFs$ make it difficult to measure and calculate with high accuracy. The real uncertainty might thus be larger than the observed scatter. 

\section{Conclusions}

New TR-LIF lifetimes were measured using two-step excitation schemes in \ion{Sc}{ii}. These measurements extend the set of available experimental values with 17 even-parity levels belonging to the excited configurations ${\rm 3d5s}$, ${\rm 3d4d}$, ${\rm 4p^2}$ and ${\rm 3d6s}$. We measured 57 $BFs$ from these upper levels using a HCL and a FTS. By combining the $BFs$ with the measured lifetimes, we derived $\log(gf)$ values from these highly-excited levels. A Hartree-Fock model that includes the main relativistic interactions along with the core-polarisation effects (HFR+CPOL) was used to determine the branching fractions and the oscillator strengths. The comparison between our HFR+CPOL and TR-LIF lifetimes along with those found in the literature \citep{buc71,arn76,pal76,vog85,mar88,ruc14,kur15} shows generally a good agreement ranging from a few percent to 20\% with the notable exceptions of the even-parity levels ${\rm 4p^2~^1D_2}$ and ${\rm 3d6s~^3D_3}$. The former discrepancy may be due to a cancellation effect that lengthens the HFR+CPOL lifetime. Owing to the good agreement ($\sim$20\%) obtained with the experimental branching fractions of \citet{law89} for low-excitation levels and ours for high-excitation levels, the HFR+CPOL branching fractions were combined with our TR-LIF lifetimes and the experimental values of \citet{mar88} to obtain rescaled semi-empirical oscillator strengths for all the 380 E1 transitions depopulating the 34 fine-structure levels for which TR-LIF lifetimes are available. This new set of oscillator
strengths were compared to the parametric calculation of \citet{ruc14} and to the Hartree-Fock values of \citet{kur15}. In both cases, the mean scatters were $\sim$20\% for the strong lines ($\log(gf)\gtrsim -1$) giving an estimate of the accuracy for these radiative parameters. Finally, the solar abundance of scandium was estimated to ${\rm log}~\epsilon_{\sun}=3.04\pm0.13$ using our rescaled semi-empirical oscillator strengths to correct the values determined in the recent study of \citet{sco15}. This value is in improved agreement with the meteoritic value (${\rm log}~\epsilon_{\text{met}}=3.05\pm0.02$) of \citet{lod09}.

\section*{Acknowledgements}

This work was financially supported by the Integrated Initiative of Infrastructure Project LASERLAB-EUROPE, contract LLC002130, and the Belgian FRS-FNRS. PQ and PP are, respectively Research Director and Research Associate of the FRS-FNRS. We acknowledge the support from the Swedish Research Council through a Linnaeus grant to the Lund Laser Centre and through project grant 2016-04185, as well as the Knut and Alice Wallenberg Foundation. MTB, JCP, and CC thank the STFC (UK) for support of their Laboratory Astrophysics research at Imperial College London.
VF is currently a post-doctoral researcher of the Return Grant programme of the Belgian Scientific Policy (BELSPO). The Belgian team is grateful to the Swedish colleagues for the warm hospitality enjoyed at the Lund Laser Centre during the two campaigns of measurements performed in June and August 2015.







\newpage

\newpage

\newpage

\newpage

\newpage

\newpage

\newpage
\begin{landscape}
\begin{table}
\label{schema}
\begin{small}
\begin{center}
\caption{Measured \ion{Sc}{ii} levels and the corresponding two$-$step excitation schemes}
\begin{tabular}{@{}lrrrrrcrr}
\hline
\multicolumn{1}{c}{Final} & \multicolumn{3}{c}{First Step Excitation} &
\multicolumn{3}{c}{Second Step Excitation} &
\multicolumn{2}{c}{Detection}  \\
\cline{2-4} \cline{5-7} \cline{8-9}
\multicolumn{1}{c}{Level$^a$} & \multicolumn{1}{c}{Starting} & 
\multicolumn{1}{c}{Intermediate} & \multicolumn{1}{c}{$\lambda_{air}$} &
\multicolumn{1}{c}{Final Level$^a$} & \multicolumn{1}{c}{$\lambda_{air}$} &
\multicolumn{1}{c}{Scheme$^b$}
 & \multicolumn{1}{c}{Lower Level$^a$} & \multicolumn{1}{c}{$\lambda_{air}$}  \\
 & \multicolumn{1}{c}{Level$^a$ (cm$^{-1}$)} & \multicolumn{1}{c}{Level$^a$ (cm$^{-1}$)} &
 \multicolumn{1}{c}{(nm)} & \multicolumn{1}{c}{(cm$^{-1}$)} & \multicolumn{1}{c}{(nm)} &
 & \multicolumn{1}{c}{(cm$^{-1}$)} & \multicolumn{1}{c}{(nm)}   \\
\hline

5s ${}^3$D$_3$ & 67.72 & 27602.45 & 363.07 & 57743.92 & 331.67 & 2$\omega$ & 27841.35 & 334.32  \\        
\\                 
5s ${}^1$D$_2$ & 67.72 & 27602.45 & 363.07 & 58252.09 & 326.17 & 2$\omega$ & 32349.98 & 385.96  \\
\\
4d ${}^1$F$_3$ & 177.76 & 29823.93 & 337.22 & 59528.42 & 336.55 & 2$\omega$ & 26081.34 & 298.89  \\
\\
4d ${}^3$D$_1$ & 177.76 & 29823.93 & 337.22 & 59875.08 & 332.67 & 2$\omega$ & 27917.78 & 312.83 \\
\\
4d ${}^3$D$_2$ & 177.76 & 29823.93 & 337.22 & 59929.46 & 332.07 & 2$\omega$ & 28021.29 & 313.31 \\
\\
4d ${}^3$D$_3$ & 177.76 & 29823.93 & 337.22 & 60001.91 & 331.27 & 2$\omega$ & 28161.17 & 313.97  \\
\\
4d ${}^3$G$_3$ & 177.76 & 29823.93 & 337.22 & 60267.16 & 328.39 & 2$\omega$ & 27443.71 & 304.57 \\ 
\\
4d ${}^1$P$_1$ & 177.76 & 29823.93 & 337.22 & 60400.41 & 326.95 & 2$\omega$ & 26081.34 & 291.30  \\
\\
\multirow{2}{*}{4d ${}^3$S$_1$ }&\multirow{2}{*}{177.76}&\multirow{2}{*}{ 29823.93} &\multirow{2}{*}{ 337.22} & \multirow{2}{*}{61071.43} &\multirow{2}{*}{ 319.93} & \multirow{2}{*}{2$\omega$ }& 29823.93 & 319.93 \\

&&  &  &  & & & 39345.52 & 460.15  \\
\\
4d ${}^3$F$_2$ & 2540.95 & 32349.98 & 335.37 & 63374.63 & 322.23 & 2$\omega$ & 27917.78 & 281.95  \\ 
\\
4d ${}^3$F$_4$ & 2540.95 & 32349.98 & 335.37 & 63528.54 & 320.64 & 2$\omega$ & 28161.17 & 282.66 \\ 
\\
\multirow{2}{*}{4d ${}^1$D$_2$ }&\multirow{2}{*}{2540.95}&\multirow{2}{*}{ 32349.98} &\multirow{2}{*}{ 335.37} & \multirow{2}{*}{64366.68} &\multirow{2}{*}{ 312.25} & \multirow{2}{*}{2$\omega$ }& 26081.34 & 261.12  \\
&&  &  &  & & & 30815.70 & 297.97  \\
\\
4d ${}^3$P$_2$ & 2540.95 & 32349.98 & 335.37 & 64705.89 & 308.98 & 2$\omega$ & 29823.93 & 286.60  \\
\\
4p${^2}$ ${}^1$D$_2$ & 177.76 & 28161.17 & 357.25 & 74433.30 & 216.04 & 3$\omega$ & 32349.98 & 237.55  \\
\\
4p${^2}$ ${}^3$P$_1$ & 177.76 & 29823.93 & 337.22 & 76360.80 & 214.82 & 3$\omega$ & 39345.52 & 270.08  \\ 
\\
\multirow{3}{*}{4p${^2}$ ${}^3$P$_2$ }&\multirow{3}{*}{177.76}&\multirow{3}{*}{ 29823.93} &\multirow{3}{*}{ 337.22} & \multirow{3}{*}{76589.30} &\multirow{3}{*}{ 213.76} & \multirow{3}{*}{3$\omega$ }& 28161.17 & 206.43  \\
&&  &  &  & & & 39115.04 & 266.77  \\
&&  &  &  & & & 39345.52 & 268.42  \\
\\
6s ${}^3$D$_3$ & 177.76 & 29823.93 & 337.22 & 77387.17 & 210.18 & 3$\omega$ & 28161.17 & 203.08  \\
\\
\hline                  
\end{tabular}
\end{center}
$^a$ All energy level values and wavelength values are from \citet{joh80}.\\
$^b$ $2\omega$ and $3\omega$ stand for, respectively, frequency doubling and
tripling excitation schemes. All first step levels are excited using a frequency doubling scheme (2$\omega$)
\end{small}
\end{table}
\end{landscape}

\newpage
\begin{landscape}

\begin{table}
\caption{\label{life}Comparison of radiative lifetimes ($\tau$) in \ion{Sc}{ii}}
\begin{small}
\begin{center}
\begin{tabular}{@{}lrrrrr}
\hline
\multicolumn{1}{c}{Level$^a$}	&	\multicolumn{1}{c}{$E^a$}	&	\multicolumn{1}{c}{$\tau_{\text{this~cal}}^b$}	&	\multicolumn{1}{c}{$\tau_{\text{this~exp}}^c$}	&	\multicolumn{1}{c}{$\tau_{\text{other~exp}}$}
&	\multicolumn{1}{c}{$\tau_{\text{other~cal}}$}	\\
 & \multicolumn{1}{c}{(cm$^{-1}$)}&\multicolumn{1}{c}{(ns)}&\multicolumn{1}{c}{(ns)}& \multicolumn{1}{c}{(ns)}
 & \multicolumn{1}{c}{(ns)}\\
\hline
${\rm 3d 4p~^1D^o_2}$	&	26081.34	&	6.65	&		&	7.5$\pm$0.4$^d$	& 6.54$^i$\\
&	&	&		&	7.16$\pm$0.18$^e$	& 7.79$^j$ \\
&	&	&		&	7.8$\pm$0.8$^h$	&\\
${\rm 3d 4p~^3F^o_2}$	&	27443.71	&	5.68	&		&	6.2$\pm$0.3$^d$	& 5.38$^i$\\
&	&	&		&	6.2$\pm$0.2$^f$	& 5.90$^j$ \\
&	&	&		&	6.5$\pm$0.4$^g$	&\\
${\rm 3d 4p~^3F^o_3}$	&	27602.45	&	5.62	&		&	6.1$\pm$0.3$^d$	& 5.32$^i$\\
&	&	&		&	& 5.83$^j$ \\
${\rm 3d 4p~^3F^o_4}$	&	27841.35	&	5.54	&		&	6.1$\pm$0.3$^d$	& 5.24$^i$\\
&	&	&		&	5.6$\pm$0.6$^h$	& 5.75$^j$\\
${\rm 3d 4p~^3D^o_1}$	&	27917.78	&	4.44	&		&	4.7$\pm$0.2$^d$	& 4.20$^i$\\
&	&	&		&	4.61$\pm$0.10$^e$	& 4.67$^j$\\
${\rm 3d 4p~^3D^o_2}$	&	28021.29	&	4.41	&		&	4.7$\pm$0.2$^d$	& 4.17$^i$\\
&	&	&		&	4.66$\pm$0.14$^e$	& 4.64$^j$ \\
${\rm 3d 4p~^3D^o_3}$	&	28161.17	&	4.38	&		&	4.7$\pm$0.2$^d$	& 4.15$^i$ \\
&	&	&		&	4.55$\pm$0.15$^e$	& 4.59$^j$\\
&	&	&		&	6.1$\pm$0.6$^h$	&\\
${\rm 3d 4p~^3P^o_0}$	&	29736.27	&	6.36	&		&	7.7$\pm$0.4$^d$	& 6.80$^i$\\
&	&	&		&	7.48$\pm$0.18$^e$	& 7.44$^j$\\
${\rm 3d 4p~^3P^o_1}$	&	29742.16	&	6.39	&		&	7.6$\pm$0.4$^d$	& 6.76$^i$\\
&	&	&		&	7.3$\pm$0.3$^e$	& 7.45$^j$\\
${\rm 3d 4p~^3P^o_2}$	&	29823.93	&	6.30	&		&	7.4$\pm$0.4$^d$	& 6.67$^i$\\
&	&	&		&	7.30$\pm$0.16$^e$	& 7.50$^j$\\
${\rm 3d 4p~^1P^o_1}$	&	30815.70	&	8.10	&		&	8.8$\pm$0.4$^d$	& 7.35$^i$  \\
&	&	&		&	8.5$\pm$0.6$^g$	& 8.76$^j$ \\
&	&	&		&	5.5$\pm$0.5$^h$	&\\
${\rm 3d 4p~^1F^o_3}$	&	32349.98	&	4.68	&		&	5.1$\pm$0.3$^d$	& 4.46$^i$ \\
&	&	&		&	5.2$\pm$0.2$^e$	& 5.20$^j$ \\
&	&	&		&	6.8$\pm$0.6$^h$	&\\
${\rm 4s 4p~^3P^o_0}$	&	39002.20	&	3.69	&		&	3.7$\pm$0.2$^d$	& 3.36$^i$ \\
&	&	&		&	& 3.66$^j$ \\
${\rm 4s 4p~^3P^o_1}$	&	39115.04	&	3.69	&		&	3.7$\pm$0.2$^d$	& 3.37$^i$ \\
&	&	&		&	& 3.67$^j$ \\
${\rm 4s 4p~^3P^o_2}$	&	39345.52	&	3.70	&		&	3.8$\pm$0.2$^d$	& 3.39$^i$\\
&	&	&		&	& 3.67$^j$ \\
${\rm 4s 4p~^1P^o_1}$	&	55715.36	&	0.88	&		&		& 0.91$^i$\\
${\rm 3d 5s~^3D_1}$	&	57551.88	&	3.49	&		&		& 3.44$^i$ \\
${\rm 3d 5s~^3D_2}$	&	57614.40	&	3.50	&		&		& 3.44$^i$ \\
${\rm 3d 5s~^3D_3}$	&	57743.92	&	3.50	&	3.20$\pm$0.20	&		& 3.44$^i$\\
${\rm 3d 5s~^1D_2}$	&	58252.09	&	3.70	&	3.26$\pm$0.20	&		& 3.66$^j$ \\
${\rm 3d 4d~^1F_3}$	&	59528.42	&	2.69	&	2.32$\pm$0.15	&		& 2.51$^i$ \\
&	&	&		&	& 2.51$^j$ \\
\hline
\end{tabular}
\end{center}
\end{small}
\end{table}
\end{landscape}

\newpage
\begin{landscape}
\addtocounter{table}{-1}
\begin{table}
\caption{Continued.}
\begin{small}
\begin{center}
\begin{tabular}{@{}lrrrrr}
\hline
\multicolumn{1}{c}{Level$^a$}	&	\multicolumn{1}{c}{$E^a$}	&	\multicolumn{1}{c}{$\tau_{\text{this~cal}}^b$}	&	\multicolumn{1}{c}{$\tau_{\text{this~exp}}^c$}	&	\multicolumn{1}{c}{$\tau_{\text{other~exp}}$}
&	\multicolumn{1}{c}{$\tau_{\text{other~cal}}$}	\\
 & \multicolumn{1}{c}{(cm$^{-1}$)}&\multicolumn{1}{c}{(ns)}&\multicolumn{1}{c}{(ns)}& \multicolumn{1}{c}{(ns)}
 & \multicolumn{1}{c}{(ns)}\\
\hline
${\rm 3d 4d~^3D_1}$	&	59875.08	&	2.72	&	2.23$\pm$0.15	&		& 2.62$^i$\\
${\rm 3d 4d~^3D_2}$	&	59929.46	&	2.74	&	2.32$\pm$0.15	&		& 2.63$^i$ \\
&	&	&		&	& 2.58$^j$ \\
${\rm 3d 4d~^3D_3}$	&	60001.91	&	2.76	&	2.41$\pm$0.20	&		& 2.65$^i$ \\
${\rm 3d 4d~^3G_3}$	&	60267.16	&	2.50	&	2.19$\pm$0.15	&	2.5$\pm$0.2$^d$	& 2.33$^i$\\
&	&	&		&	& 2.47$^j$ \\
${\rm 3d 4d~^3G_4}$	&	60348.46	&	2.52	&		&	2.4$\pm$0.2$^d$	& 2.35$^i$ \\
&	&	&		&	& 2.49$^j$ \\
${\rm 3d 4d~^1P_1}$	&	60400.41	&	2.89	&	2.44$\pm$0.15	&		& 2.69$^i$ \\
&	&	&		&	& 2.63$^j$ \\
${\rm 3d 4d~^3G_5}$	&	60457.12	&	2.54	&		&	2.5$\pm$0.2$^d$	& 2.38$^i$ \\
&	&	&		&	& 2.51$^j$ \\
${\rm 3d 4d~^3S_1}$	&	61071.43	&	2.82	&	2.45$\pm$0.15	&		& 2.77$^i$ \\
&	&	&		&	& 2.78$^j$ \\
${\rm 3d 4d~^3F_2}$	&	63374.63	&	2.40	&	2.15$\pm$0.10	&		& 2.05$^i$\\
&	&	&		&	& 2.43$^j$ \\
${\rm 3d 4d~^3F_3}$	&	63445.16	&	2.41	&		&		& 2.05$^i$ \\
${\rm 3d 4d~^3F_4}$	&	63528.54	&	2.43	&	2.19$\pm$0.10	&		&2.07$^i$\\
&	&	&		&	& 2.47$^j$ \\
${\rm 3d 4d~^1D_2}$	&	64366.68	&	2.73	&	2.25$\pm$0.15	&		& 2.26$^i$ \\
${\rm 3d 4d~^3P_0}$	&	64615.77	&	3.21	&		&		& 2.65$^i$ \\
${\rm 3d 4d~^3P_1}$	&	64646.70	&	3.21	&		&		& 2.65$^i$ \\
${\rm 3d 4d~^3P_2}$	&	64705.89	&	3.19	&	2.51$\pm$0.15	&		& 2.65$^i$ \\
${\rm 3d 4d~^1G_4}$	&	65236.04	&	3.17	&		&		& 2.45$^i$ \\
${\rm 3d 4d~^1S_0}$	&	67216.56	&	3.87	&		&		& 2.74$^i$ \\
${\rm 4p^2~^1D_2}$	&	74433.30	&	5.96	&	3.80$\pm$0.15	&		& 6.80$^i$ \\
${\rm 4p^2~^3P_0}$	&	76243.20	&	1.17	&		&		& 1.28$^i$ \\
${\rm 4p^2~^3P_1}$	&	76360.80	&	1.17	&	1.14$\pm$0.06	&		&1.28$^i$ \\
${\rm 4p^2~^3P_2}$	&	76589.30	&	1.18	&	1.09$\pm$0.06	&		& 1.30$^i$ \\
&	&	&		&	& 1.03$^j$ \\
${\rm 3d 6s~^3D_1}$ &	77195.19	&	5.56	&		&		& 5.00$^i$ \\
${\rm 3d 6s~^3D_2}$	&	77256.99	&	5.55	&		&		& 5.00$^i$\\
${\rm 3d 6s~^3D_3}$	&	77387.17	&	5.54	&	3.73$\pm$0.25	&		& 4.98$^i$\\
${\rm 3d 6s~^1D_2}$	&	77833.88	&	6.61	&		&		&  6.94$^i$\\
\hline
\end{tabular}
\end{center}
$^a$  \citet{joh80}.\\
$^b$ HFR+CPOL calculation. This work.\\
$^c$ TR-LIF measurements. This work.\\
$^d$ TR-LIF measurements by \citet{mar88}.\\
$^e$ TR-LIF measurements by \citet{vog85}.\\
$^f$ TR-LIF measurements by \citet{arn76}.\\
$^g$ Beam-foil spectroscopy by \citet{pal76}.\\
$^h$ Beam-foil spectroscopy by \citet{buc71}.\\
$^i$ Hartree-Fock calculation by \citet{kur15}.\\
$^j$ Parametric method calculation by \citet{ruc14}.
\end{small}
\end{table}
\end{landscape}

\newpage

\begin{landscape}
\begin{table}
\caption{Presentation of experimental $\log(gf)$ values together with the transition, wavelength, $\lambda$, wavenumber, $\sigma$, measured branching fraction, $BF_{\text{exp}}$, experimental transition probability, $A_\text{exp}$ , and the corresponding rescaled semi-empirical $\log(gf)$ values of this work. The radiative lifetimes, $\tau$, are TR-LIF measurements from this work.}
\label{exp}
\begin{small}
\begin{center}
\begin{tabular}{@{}lclccccccrrr}
\hline
 \multicolumn{2}{c}{Upper level$^a$ } &  \multicolumn{2}{c}{Lower level$^a$} & $\lambda_{\text{exp}}^a$ & $\sigma_{\text{exp}}^a$ & $\sigma_{\text{theo}}^b$& $BF_{\text{exp}}$ & $BF$ unc. & $A_\text{exp}$ & $\log(gf) $ & $\log(gf)_{\text{resc}}$ \\
 Config. & Energy (cm$^{-1}$) & Config. & Energy (cm$^{-1}$) & (nm) & (cm$^{-1}$) & (cm$^{-1}$) & & $\%$ & (s$^{-1}$) & Exp. & Calc. \\
\hline
3d5s $^3$D$_3$ & 57744 & 3d4p $^3$F$_3^o$ & 27602 & 331.673 & 30141.50 & 30176 & 6.21E-02 & 4 & 1.94E+07 & -0.65$\pm$0.03 & -0.75 \\
$\tau$= 3.20$\pm$0.20 ns & & 3d4p  $^3$F$_4^o$ & 27841 & 334.323 & 29902.57 & 29944 &  4.05E-01 & 3 & 1.27E+08 & 0.17$\pm$0.03 & 0.14 \\
 & & 3d4p $^3$D$_2^o$ & 28021 & 336.347 & 29722.58 & 29771 & 5.29E-02 & 4 & 1.65E+07 & -0.71$\pm$0.03 & -0.69 \\
 & & 3d4p $^3$D$_3^o$ & 28161 & 337.938 & 29582.76 & 29620 & 3.60E-01 & 3 & 1.12E+08 & 0.13$\pm$0.03 & 0.17 \\
& & 3d4p $^3$P$_2^o$ & 29824 & 358.064 & 27919.88 & 27888 & 1.20E-01 & 4 & 3.75E+07 & -0.30$\pm$0.03 & -0.28 \\
& & \textit{Residual} & & & & & 3.37E-03 & &  \\
\\
3d5s $^1$D$_2$ &58252 & 3d4p $^1$D$_2^o$ & 26081 & 310.751 & 32179.68 & 32040 & 4.90E-01 & 2 & 1.50E+08 & 0.04$\pm$0.03 & -0.09 \\
$\tau$= 3.26$\pm$0.20 ns      & & 3d4p $^3$F$_2^o$ & 27444  & 324.493 & 30808.34 & 30866 & 1.08E-02 & 16 &3.30E+06 & -1.58$\pm$0.07 & -1.64 \\
                                      & & 3d4p $^1$P$_1^o$ & 30816 & 364.376 & 27436.43 & 27508 & 1.48E-01 & 4 & 4.55E+07 & -0.34$\pm$0.03 & -0.32 \\
                                      & & 3d4p $^1$F$_3^o$ & 32350 & 385.960 & 25902.13 & 25865 & 3.51E-01 & 5 & 1.08E+08 & 0.08$\pm$0.03 & 0.18 \\
& & \textit{Residual} & & & & & 2.10E-02 & &  \\
\\                                      
3d4d $^1$F$_3$ & 59528 & 3d4p $^1$D$_2^o$ & 26081 & 298.893 & 33447.17 & 33296 & 8.22E-01 & 0.5 & 3.54E+08 & 0.52$\pm$0.03 & 0.47 \\
$\tau$= 2.32$\pm$0.15 ns     & & 3d4p $^3$D$_3^o$ & 28161 & 318.712 & 31367.21 & 31408 & 6.54E-03 & 14 & 2.82E+06 & -1.52$\pm$0.06 & -1.45 \\
                                      & & 3d4p $^1$F$_3^o$ & 32350 & 367.834 & 27178.50 & 27121 & 1.72E-01 & 7 & 7.40E+07 & 0.02$\pm$0.04 & 0.20 \\
& & \textit{Residual} & &  & & & 5.10E-03 & &  \\
\\    
3d4d $^3$D$_1$ & 59875 &  3d4p $^3$F$_2^o$ & 27444 & 308.254 & 32431.14 & 32475 & 1.22E-01 & 5 & 5.49E+07 & -0.63$\pm$0.04 & -0.63 \\
$\tau$= 2.23$\pm$0.15 ns        &  & 3d4p $^3$D$_1^o$  & 27918 & 312.827 & 31957.28 & 32026 & 4.57E-01 &  3 & 2.05E+08 & -0.04$\pm$0.03 & -0.11 \\
                                         & & 3d4p $^3$D$_2^o$ & 28021 & 313.843 & 31853.76 & 31913 &1.17E-01 & 5 & 5.24E+07 & -0.63$\pm$0.04 & -0.65 \\
                                         & & 3d4p $^3$P$_0^o$ & 29736 & 331.703 & 30138.84 & 30134 & 1.79E-01 & 4 & 8.02E+07 & -0.40$\pm$0.03 & -0.37 \\
                                         & & 3d4p $^3$P$_1^o$ & 29742 & 331.768 & 30132.91 & 30123 & 1.25E-01 & 5 & 5.61E+07 & -0.56$\pm$0.04 & -0.50\\
& & \textit{Residual} & & & & & 3.75E-02 & &  \\
\\    
3d4d $^3$D$_2$ & 59929 &  3d4p $^3$F$_3^o$ & 27602 & 309.249 & 32327.05 & 32372 & 1.21E-01 & 5 & 5.23E+07 & -0.43$\pm$0.03 & -0.45 \\
$\tau$= 2.32$\pm$0.15 ns       &  & 3d4p $^3$D$_1^o$  & 27918 & 312.296 & 32011.74 & 32081 & 8.39E-02 & 5 & 3.62E+07 & -0.58$\pm$0.03 & -0.63 \\
                                         & & 3d4p $^3$D$_2^o$ & 28021 & 313.309 & 31908.30 & 31968 & 4.12E-01 & 3 & 1.78E+08 & 0.12$\pm$0.03 & 0.06 \\
                                         & & 3d4p $^3$D$_3^o$ & 28161 & 314.688 & 31768.28 & 31816 & 6.61E-02 & 5 & 2.85E+07 & -0.67$\pm$0.03 & -0.68 \\
                                         & & 3d4p $^3$P$_1^o$ & 29742 & 331.170 & 30187.30 & 30178 & 2.50E-01 & 4 & 1.08E+08 & -0.05$\pm$0.03 & -0.02 \\
                                         & & 3d4p $^3$P$_2^o$ & 29824 & 332.069 & 30105.53 & 30084 & 6.67E-02 & 5 & 2.87E+07 & -0.62$\pm$0.03 & -0.56 \\
                                       & &   \textit{Residual} & & & & & 3.42E-02 & &  \\
\\ 
3d4d $^3$D$_3$ & 60002 & 3d4p $^3$F$_4^o$ & 27841 & 310.850 & 32160.62 & 32214 & 9.39E-02 & 6 & 3.90E+07 & -0.40$\pm$0.04 & -0.34 \\
$\tau$= 2.41$\pm$0.20 ns             &  & 3d4p $^3$D$_2^o$  & 28021 & 312.599 & 31980.37 & 32041 & 4.35E-02 & 7 & 1.81E+07 & -0.73$\pm$0.06 & -0.63 \\
                                        & & 3d4p $^3$D$_3^o$ & 28161 & 313.972 & 31840.77 & 31890 & 5.16E-01 & 4 & 2.14E+08 & 0.35$\pm$0.04 & 0.28 \\
                                        & & 3d4p $^3$P$_2^o$ & 29824 & 331.272 & 30178.03 & 30157 & 3.47E-01 & 5 & 1.44E+08 & 0.22$\pm$0.04 & 0.24 \\
                                        & &  \textit{Residual} & & & & &  3.51E-02 & &  \\
\\                                           
\hline                  
\end{tabular}
\end{center}
\end{small}
\end{table}
\end{landscape}

\newpage
\begin{landscape}
\addtocounter{table}{-1}
\begin{table}
\caption{Continued.}
\label{exp}
\begin{small}
\begin{center}
\begin{tabular}{@{}lclccccccrrr}
\hline
 \multicolumn{2}{c}{Upper level$^a$ } &  \multicolumn{2}{c}{Lower level$^a$} & $\lambda_{\text{exp}}^a$ & $\sigma_{\text{exp}}^a$ & $\sigma_{\text{theo}}^b$& $BF_{\text{exp}}$ & $BF$ unc. & $A_\text{exp}$ & $\log(gf) $ & $\log(gf)_{\text{resc}}$ \\
 Config. & Energy (cm$^{-1}$) & Config. & Energy (cm$^{-1}$) & (nm) & (cm$^{-1}$) & (cm$^{-1}$) & & $\%$ & (s$^{-1}$) & Exp. & Calc. \\
\hline
3d4d $^3$G$_3$ & 60267 & 3d4p $^3$F$_2^o$ & 27444 & 304.572 & 32823.36 & 32822 & 9.26E-01 & 1 & 4.23E+08 & 0.61$\pm$0.03 & 0.61 \\
$\tau$= 2.19$\pm$0.15 ns             &  & 3d4p $^3$F$_3^o$  & 27602 & 306.052 & 32664.51 & 32664 & 7.41E-02 & 8 & 3.38E+07 & -0.48$\pm$0.04 & -0.47 \\
                                        & & \textit{Residual} & & & & &  6.20E-03 & & & & \\
\\
3d4d $^1$P$_1$ & 60400 & 3d4p  $^1$D$_2^o$ & 26081 & 291.298 & 34319.09 & 34206 & 3.98E-01 & 7 & 1.63E+08 & -0.21$\pm$0.04 & -0.33 \\
$\tau$= 2.44$\pm$0.15 ns            &  & 3d4p $^1$P$_1^o$  & 30816 & 337.915 & 29584.65 & 29673 & 6.02E-01 & 5 & 2.47E+08 & 0.10$\pm$0.03 & 0.14 \\
                                    & &     \textit{Residual} & & & & & 7.27E-02 & & & & \\
\\
3d4d $^3$S$_1$ & 61071 & 3d4p $^3$P$_0^o$  & 29736 & 319.038 & 31335.12 & 31336 & 1.13E-01 & 7 & 4.60E+07 & -0.68$\pm$0.04 & -0.70 \\
$\tau$= 2.45$\pm$0.15 ns          &  & 3d4p $^3$P$_1^o$  & 29742 & 319.098 & 31329.24 & 31326 & 2.84E-01 & 6 & 1.16E+08 & -0.28$\pm$0.04 & -0.25 \\
                                       & & 3d4p $^3$P$_2^o$  & 29824 & 319.933 & 31247.50 & 31231 & 5.73E-01 & 4 & 2.34E+08 & 0.03$\pm$0.03 & -0.02 \\
                                       & & 3d4p $^1$P$_1^o$  & 30816 & 330.421 & 30255.76 & 30319 & 3.04E-02 & 10 & 1.24E+07 & -1.21$\pm$0.05 & -1.28 \\
                             & &            \textit{Residual} & & & & & 6.10E-02 & & & & \\
\\
3d4d $^3$F$_2$ & 63375 & 3d4p $^3$F$_2^o$ & 27444 & 278.230 & 35930.81 & 35960 & 3.57E-01 & 5 & 1.66E+08 & -0.02$\pm$0.03 & -0.07 \\
$\tau$= 2.15$\pm$0.10 ns           &  & 3d4p $^3$F$_3^o$  & 27602 & 279.464 & 35772.19 & 35802 & 3.72E-02 & 9 & 1.73E+07 & -0.99$\pm$0.04 & -0.95 \\
                                      & & 3d4p $^3$D$_1^o$  & 27918 & 281.950 & 35456.96 & 35511 & 5.27E-01 & 3 & 2.45E+08 & 0.17$\pm$0.02 & 0.17 \\
                                      & & 3d4p $^3$D$_2^o$  & 28021 & 282.776 & 35353.30 & 35398 & 7.88E-02 & 6 & 3.66E+07 & -0.66$\pm$0.03 & -0.60 \\
                                   & &      \textit{Residual} & & & & & 1.59E-02 & & & & \\
\\                                      
3d4d $^3$F$_4$ & 63529 & 3d4p $^3$F$_4^o$ & 27841 & 280.130 & 35687.12 & 35726 & 3.42E-01 & 6 & 1.56E+08 & 0.22$\pm$0.03 & 0.21 \\
$\tau$= 2.19$\pm$0.10 ns            &  & 3d4p $^3$D$_3^o$ & 28161 & 282.663 & 35367.30 & 35402 & 6.58E-01 & 3 & 3.01E+08 & 0.51$\pm$0.02 & 0.51 \\
                                    & &     \textit{Residual} & & & & & 8.18E-03 & & & & \\
\\                  
3d4d $^1$D$_2$ & 64367 & 3d4p $^1$D$_2^o$ & 26081 & 261.119 & 38285.22 & 38187 & 7.25E-01 & 4 & 3.22E+08 & 0.22$\pm$0.03 & 0.11 \\
$\tau$= 2.25$\pm$0.15 ns              &  & 3d4p $^3$F$_2^o$ & 27444 & 270.754 & 36923.00 & 37012 & 1.69E-02 & 12 & 7.51E+06 & -1.38$\pm$0.06 & -1.49 \\
                                        & & 3d4p $^3$P$_1^o$ & 29742 & 288.728 & 34624.48 & 34661 & 2.27E-02 & 16 & 1.01E+07 & -1.20$\pm$0.07 & -1.07 \\
                                        & & 3d4p $^1$P$_1^o$ & 30816 & 297.967 & 33550.90 & 33654 & 2.35E-01 & 8 & 1.05E+08 & -0.16$\pm$0.04 & -0.02 \\
                                    & &     \textit{Residual} & & & & & 4.94E-02 & & & & \\
\\
3d4d $^3$P$_2$ & 64706 & 3d4p $^3$D$_3^o$ & 28161 & 273.556 & 36544.66 & 36597 & 1.59E-01 & 8 & 6.32E+07 & -0.45$\pm$0.04 & -0.46 \\
$\tau$= 2.51$\pm$0.15 ns             &  & 3d4p $^3$P$_1^o$ & 29742 & 285.927 & 34963.68 & 34960 & 1.78E-01 & 6 & 7.09E+07 & -0.36$\pm$0.04 & -0.40 \\
                                        & & 3d4p $^3$P$_2^o$ & 29824 & 286.597 & 34881.86 & 34865 & 6.33E-01 & 3 & 2.52E+08 & 0.19$\pm$0.03 & 0.16 \\
                                        & & 3d4p $^1$P$_1^o$ & 30816 & 294.984 & 33890.19 & 33953 & 3.08E-02 & 8 & 1.23E+07 & -1.10$\pm$0.04 & -0.98 \\
                                   & &      \textit{Residual} & & & & & 4.86E-02 & & & & \\
\\                        
4p$^2$ $^3$P$_1$ & 76361 & 3d4p $^3$D$_2^o$ & 28021 & 206.804 & 48339.50 & 48384 & 3.16E-01 & 7 & 2.77E+08 & -0.27$\pm$0.04 & -0.41 \\
$\tau$= 1.14$\pm$0.06 ns                  &  & 4s4p $^3$P$_0^o$ & 39002 & 267.597 & 37358.69 & 37358 & 2.29E-01 & 6 & 2.01E+08 & -0.19$\pm$0.03 & -0.18 \\
                                           & & 4s4p $^3$P$_1^o$ & 39115 & 268.407 & 37245.53 & 37245 & 1.80E-01 & 6 & 1.58E+08 & -0.29$\pm$0.03 & -0.31 \\
                                           & & 4s4p $^3$P$_2^o$ & 39346 & 270.079 & 37014.70 & 37014 & 2.75E-01 & 6 & 2.41E+08 & -0.10$\pm$0.03 & -0.09 \\
                                         & & \textit{Residual} & & & & & 6.46E-02 & & & & \\
\\
\hline                  
\end{tabular}
\end{center}
\end{small}
\end{table}
\end{landscape}

\newpage
\begin{landscape}
\addtocounter{table}{-1}
\begin{table}
\caption{Continued.}
\label{exp}
\begin{small}
\begin{center}
\begin{tabular}{@{}lclccccccrrr}
\hline
 \multicolumn{2}{c}{Upper level$^a$ } &  \multicolumn{2}{c}{Lower level$^a$} & $\lambda_{\text{exp}}^a$ & $\sigma_{\text{exp}}^a$ & $\sigma_{\text{theo}}^b$& $BF_{\text{exp}}$ & $BF$ unc. & $A_\text{exp}$ & $\log(gf) $ & $\log(gf)_{\text{resc}}$ \\
 Config. & Energy (cm$^{-1}$) & Config. & Energy (cm$^{-1}$) & (nm) & (cm$^{-1}$) & (cm$^{-1}$) & & $\%$ & (s$^{-1}$) & Exp. & Calc. \\
\hline

4p$^2$ $^3$P$_2$ & 76589 & 3d4p $^3$D$_2^o$ & 28021 & 205.831 & 48568.03 & 48615 & 5.63E-02 & 27 & 5.17E+07 & -0.78$\pm$0.11 & -0.88 \\
$\tau$= 1.09$\pm$0.06 ns                 &  & 3d4p $^3$D$_3^o$ & 28161 & 206.426 & 48428.15 & 48464 & 2.95E-01 & 6 & 2.71E+08 & -0.06$\pm$0.03 & -0.13 \\
                                           & & 4s4p $^3$P$_1^o$ & 39115 & 266.771 & 37474.35 & 37477 & 1.65E-01 & 6 & 1.52E+08 & -0.09$\pm$0.03 & -0.06 \\
                                           & & 4s4p $^3$P$_2^o$ & 39346 & 268.422 & 37243.72 & 37246 & 4.83E-01 & 4 & 4.43E+08 & 0.38$\pm$0.03 & 0.41 \\
                                         & & \textit{Residual} & & & & & 1.18E-02 & & & & \\
\\                                     
\hline                  
\end{tabular}
\end{center}
\end{small}
\begin{small}
$^a$ Energy level, wavelength, and wavenumber values are taken from \citet{joh80} which are available in NIST database \citep{kra15}.
\\
$^b$ Theoretical wavenumber values are from the calculations of this work.
\end{small}

\end{table}
\end{landscape}

\newpage
\begin{table}
\caption{\label{epar} Radial parameters adopted in the HFR+CPOL calculations
for the even-parity configurations of \ion{Sc}{ii}. The parameters not listed here
have been fixed to their $ab~initio$ values or to 80\% of their HFR+CPOL values 
for the electrostatic integrals.}
\begin{center}
\begin{tabular}{@{}llrrrl}
\hline
\multicolumn{1}{c}{Config.}  & \multicolumn{1}{c}{Parameter} & \multicolumn{1}{c}{$Ab~initio$} &
 \multicolumn{1}{c}{Fitted} & \multicolumn{1}{c}{Ratio} & \multicolumn{1}{c}{Note$^a$} \\
 & & \multicolumn{1}{c}{(cm$^{-1}$)} & \multicolumn{1}{c}{(cm$^{-1}$)} & & \\
\hline
${\rm 3d 4s}$ & $E_{av}$ & 1075 & 1137 & & \\
 & $\zeta_{3d}$ & 83  & 72 & 0.87 & \\
 & $G^2(3d4s)$ & 11351 & 9883 & 0.87 & \\
${\rm 3d 5s}$ & $E_{av}$ & 57881 & 58144 & & \\
 & $\zeta_{3d}$ & 87 & 79 & 0.91 & \\
 & $G^2(3d5s)$ & 2071 & 1851 & 0.89 & \\
${\rm 3d 6s}$ & $E_{av}$ & 77397 & 77497 & & \\
 & $\zeta_{3d}$ & 88 & 82 & 0.93 & \\
 & $G^2(3d6s)$ & 789 & 631 & 0.80 & F \\
${\rm 3d 7s}$ & $E_{av}$ & 86487 & 86549 & & \\
 & $\zeta_{3d}$ & 88 & 69 & 0.78 & \\
 & $G^2(3d7s)$ & 393 & 314 & 0.80 & F \\
${\rm 3d^2}$ & $E_{av}$ & 11721 & 9531 & & \\
 & $F^2(3d3d)$ & 49657  & 37346 & 0.75 & \\
 & $F^4(3d3d)$ & 30556 & 22011 & 0.72 \\
 &$\alpha$ & 0 & 64 & & \\
 &$\beta$ & 0 & -962 & & \\
 &$T$ & 0 & 3 & & \\
 & $\zeta_{3d}$ & 65 & 59 & 0.91 & \\
${\rm 3d 4d}$ & $E_{av}$ & 62210 & 62852 & & \\
 & $\zeta_{3d}$ & 87 & 79 & 0.91 & \\
 & $\zeta_{4d}$ & 8 & 8 & 1.00 & F \\
 & $F^2(3d4d)$ & 7539 & 5977 & 0.79  & \\
 & $F^4(3d4d)$ & 3599 & 2816 & 0.78 & \\
 & $G^0(3d4d)$ & 6862 & 2467 & 0.36 & \\
 & $G^2(3d4d)$ & 4352 & 3238 & 0.74  & \\
 & $G^4(3d4d)$ & 2927  & 2327 & 0.80 & \\
${\rm 3d 5d}$ & $E_{av}$ & 79393 & 79170 & & \\
 & $\zeta_{3d}$ & 87 & 86 & 0.99 & \\
 & $\zeta_{5d}$ & 3 & 3 & 1.00 & F \\
 & $F^2(3d5d)$ & 2896 & 2158 & 0.75  & \\
 & $F^4(3d5d)$ & 1388 & 1099 & 0.79 & \\
 & $G^0(3d5d)$ & 2416 & 1008 & 0.42  & R \\
 & $G^2(3d5d)$ & 1640 & 684 & 0.42 & R \\
 & $G^4(3d5d)$ & 1122 & 469 & 0.42 & R \\
${\rm 3d 6d}$ & $E_{av}$ & 87550 & 87894 & & \\
 & $\zeta_{3d}$ & 88 & 88 & 1.00 & F\\
 & $\zeta_{6d}$ & 2 &  2 & 1.00 & F \\
 & $F^2(3d6d)$ & 1458 & 1166& 0.80 & F\\
 & $F^4(3d6d)$ &705 & 564 & 0.80 & F \\
 & $G^0(3d6d)$ & 1176 & 941 & 0.80 & F \\
 & $G^2(3d6d)$ & 822 & 658 & 0.80 & F \\
 & $G^4(3d6d)$ & 571 & 454 & 0.80 & F \\
${\rm 3d 5g}$ & $E_{av}$ & 85492 & 85761 & & \\
 & $\zeta_{3d}$ & 88 & 78  & 0.89 & \\
 & $\zeta_{5g}$ & 0 & 0 & 1.00 & F \\
 & $F^2(3d5g)$ & 465 & 420  & 0.90  & \\
 & $F^4(3d5g)$ & 42 & 34 & 0.81 & \\
 & $G^2(3d5g)$ & 6 & 5 & 0.80 & F \\
 & $G^4(3d5g)$ & 4 & 3 & 0.80  & F \\
 & $G^6(3d5g)$ & 2 & 2 & 0.80 & F \\
\hline
\end{tabular}
\end{center}
\end{table}

\newpage
\addtocounter{table}{-1}
\begin{table}
\caption{Continued.}
\begin{center}
\begin{tabular}{@{}llrrrl}
\hline
\multicolumn{1}{c}{Config.}  & \multicolumn{1}{c}{Parameter} & \multicolumn{1}{c}{$Ab~initio$} &
 \multicolumn{1}{c}{Fitted} & \multicolumn{1}{c}{Ratio} & \multicolumn{1}{c}{Note$^a$} \\
 & & \multicolumn{1}{c}{(cm$^{-1}$)} & \multicolumn{1}{c}{(cm$^{-1}$)} & & \\
\hline
${\rm 4s^2}$ & $E_{av}$ & 16845 & 16876 & & \\
${\rm 4s 5s}$ & $E_{av}$ &78974 & 79141 & & \\
 & $G^0(4s5s)$ & 2341 & 1765 & 0.75  & \\
${\rm 4s 4d}$ & $E_{av}$ & 83034 & 82930 & & \\
 & $\zeta_{4d}$ & 9 & 9 & 1.00 & F \\
 & $G^2(4s4d)$ & 6830 & 5464 & 0.80 & F \\ 
${\rm 4p^2}$ &  $E_{av}$ & 77789 & 80625 & & \\
 & $F^2(4p4p)$ &28516  & 29802 & 1.05 & \\
 & $\zeta_{4p}$ & 199 & 253 & 1.27 & \\ 
\hline
\end{tabular}
\end{center}
$^a$ F and R stand for, respectively, a fixed parameter value and a fixed ratio between
these parameters.
\end{table}

\newpage
\begin{table}
\caption{\label{opar} Radial parameters adopted in the HFR+CPOL calculations
for the odd-parity configurations of \ion{Sc}{ii}. The parameters not listed here
have been fixed to their $ab~initio$ values or to 80\% of their HFR+CPOL values 
for the electrostatic integrals.}
\begin{center}
\begin{tabular}{@{}llrrrl}
\hline
\multicolumn{1}{c}{Config.}  & \multicolumn{1}{c}{Parameter} & \multicolumn{1}{c}{$Ab~initio$} &
 \multicolumn{1}{c}{Fitted} & \multicolumn{1}{c}{Ratio} & \multicolumn{1}{c}{Note$^a$} \\
 & & \multicolumn{1}{c}{(cm$^{-1}$)} & \multicolumn{1}{c}{(cm$^{-1}$)} & & \\
\hline
${\rm 3d 4p}$ & $E_{av}$ & 28207 & 28996 & & \\
 & $\zeta_{3d}$ & 85  & 91 & 1.07  & \\
 & $\zeta_{4p}$ & 146  & 162 & 1.11 & \\
 & $F^2(3d4p)$ & 14647 & 12024 & 0.82  & \\
 & $G^1(3d4p)$ & 6709 & 6289 & 0.94  & \\
 & $G^3(3d4p)$ & 5361 & 4338 & 0.81  & \\ 
${\rm 3d 5p}$ & $E_{av}$ & 66759 & 66915 & & \\
 & $\zeta_{3d}$ & 87  & 73 & 0.84 & \\
 & $\zeta_{5p}$ & 50   & 50 & 1.00  & F \\
 & $F^2(3d5p)$ & 4168  & 3375 & 0.81 & \\
 & $G^1(3d5p)$ & 1560 & 1397 & 0.90 & \\
 & $G^3(3d5p)$ & 1380 & 900 & 0.65  & \\
${\rm 3d 4f}$ & $E_{av}$ & 75021 & 75609 & & \\
 & $\zeta_{3d}$ & 88  & 74 & 0.84 & \\
 & $\zeta_{4f}$ & 0  & 0 & 1.00  & F \\
 & $F^2(3d4f)$ & 2127 & 1766 & 0.83 & \\
 & $F^4(3d4f)$ & 514 & 367 & 0.71 & \\
 & $G^1(3d4f)$ & 420 & 354 & 0.84 & \\
 & $G^3(3d4f)$ & 242 & 194 & 0.80 & F \\ 
 & $G^5(3d4f)$ & 166 & 133 & 0.80 & F \\
${\rm 3d 5f}$ & $E_{av}$ & 85220 & 85564 & & \\
 & $\zeta_{3d}$ &  88 & 91 & 1.03 & \\
 & $\zeta_{5f}$ & 0  & 0 & 1.00 & F \\
 & $F^2(3d5f)$ & 1051 & 841 &  0.80  & F \\
 & $F^4(3d5f)$ & 296 & 238 & 0.80  & F \\
 & $G^1(3d5f)$ & 289 & 232 & 0.80 & F \\
 & $G^3(3d5f)$ & 168 & 135 & 0.80 & F \\ 
 & $G^5(3d5f)$ & 116 & 93 & 0.80 & F \\
${\rm 3d 6f}$ & $E_{av}$ & 90728 & 91031 & & \\
 & $\zeta_{3d}$ & 88  & 88 & 1.00  & F \\
 & $\zeta_{6f}$ & 0  & 0 & 1.00 & F \\
 & $F^2(3d6f)$ & 597 & 478 & 0.80  & F \\
 & $F^4(3d6f)$ & 181 & 145 & 0.80  & F \\
 & $G^1(3d6f)$ & 188 & 151 & 0.80 & F \\
 & $G^3(3d6f)$ & 111 & 88 & 0.80  & F \\ 
 & $G^5(3d6f)$ & 76 & 61 & 0.80 & F \\
${\rm 4s 4p}$ & $E_{av}$ & 41287 & 43719 & & \\
 & $\zeta_{4p}$ & 197 & 242 & 1.23 &  \\
 & $G^1(4s4p)$ & 37326 & 27686 & 0.74  &  \\
\hline
\end{tabular}
\end{center}
$^a$ F stands for a fixed parameter value.
\end{table}

\newpage
\begin{landscape}
\begin{table}
\caption{Calculated branching fractions ($BF$), oscillator strengths ($\log(gf)$) and transition probabilities ($gA$) along with the corresponding scaled values ($\log(gf)_{\text{resc}}$, $gA_{\text{resc}}$) in \ion{Sc}{ii}.
Only the transitions depopulating the levels for which the lifetime has been measured are listed. The experimental lifetimes ($\tau_u$) used to scale the $f$- and $A$-values are also reported.}
\begin{center}
\begin{small}
\begin{tabular}{rrrrrrrrrrrrrr}
\hline
\multicolumn{3}{c}{Upper level$^a$} & \multicolumn{1}{c}{$\tau_u$ (ns)} & \multicolumn{3}{c}{Lower level$^a$}&
 \multicolumn{1}{c}{$\lambda^b$ (nm)} & \multicolumn{1}{c}{$BF$} & \multicolumn{1}{c}{$gA$ (s$^{-1}$)} & \multicolumn{1}{c}{$gA_{\text{resc}}$ (s$^{-1}$)} & \multicolumn{1}{c}{$\log(gf)$} & \multicolumn{1}{c}{$\log(gf)_{\text{resc}}$} & \multicolumn{1}{c}{CF$^c$} \\
\hline
26081	&	(o)	&	2	&	7.5$^d$	&	0	&	(e)	&	1	&	383.307	&	5.91E-03	&	4.44E+06	&	3.94E+06	&	-2.01	&	-2.06	&	0.468	\\
	&		&		&		&	68	&	(e)	&	2	&	384.305	&	1.40E-02	&	1.05E+07	&	9.32E+06	&	-1.64	&	-1.69	&	0.927	\\
	&		&		&		&	178	&	(e)	&	3	&	385.938	&	1.65E-05	&	1.24E+04	&	1.10E+04	&	-4.56	&	-4.61	&	0.006	\\
	&		&		&		&	2541	&	(e)	&	2	&	424.682	&	9.73E-01	&	7.31E+08	&	6.49E+08	&	0.29	&	0.24	&	0.975	\\
	&		&		&		&	4803	&	(e)	&	2	&	469.827	&	7.55E-04	&	5.67E+05	&	5.03E+05	&	-2.73	&	-2.78	&	0.260	\\
	&		&		&		&	4884	&	(e)	&	3	&	471.616	&	3.77E-05	&	2.83E+04	&	2.51E+04	&	-4.03	&	-4.08	&	0.047	\\
	&		&		&		&	10945	&	(e)	&	2	&	660.460	&	6.19E-03	&	4.65E+06	&	4.13E+06	&	-1.53	&	-1.57	&	0.036	\\
	&		&		&		&	12102	&	(e)	&	1	&	715.119	&	6.88E-07	&	5.17E+02	&	4.59E+02	&	-5.41	&	-5.45	&	0.009	\\
	&		&		&		&	12154	&	(e)	&	2	&	717.836	&	8.89E-06	&	6.68E+03	&	5.93E+03	&	-4.30	&	-4.34	&	0.008	\\
27444	&	(o)	&	2	&	6.2$^d$	&	0	&	(e)	&	1	&	364.278	&	7.27E-01	&	6.40E+08	&	5.87E+08	&	0.11	&	0.07	&	0.888	\\
	&		&		&		&	68	&	(e)	&	2	&	365.180	&	1.57E-01	&	1.38E+08	&	1.26E+08	&	-0.56	&	-0.60	&	0.894	\\
	&		&		&		&	178	&	(e)	&	3	&	366.653	&	6.61E-03	&	5.82E+06	&	5.33E+06	&	-1.93	&	-1.97	&	0.927	\\
	&		&		&		&	2541	&	(e)	&	2	&	401.448	&	1.27E-02	&	1.12E+07	&	1.03E+07	&	-1.57	&	-1.61	&	0.837	\\
	&		&		&		&	4803	&	(e)	&	2	&	441.556	&	8.94E-02	&	7.87E+07	&	7.21E+07	&	-0.64	&	-0.68	&	0.971	\\
	&		&		&		&	4884	&	(e)	&	3	&	443.135	&	7.01E-03	&	6.17E+06	&	5.65E+06	&	-1.74	&	-1.78	&	0.466	\\
	&		&		&		&	10945	&	(e)	&	2	&	605.924	&	1.13E-04	&	9.98E+04	&	9.15E+04	&	-3.26	&	-3.30	&	0.046	\\
	&		&		&		&	12102	&	(e)	&	1	&	651.617	&	2.08E-05	&	1.83E+04	&	1.68E+04	&	-3.93	&	-3.97	&	0.676	\\
	&		&		&		&	12154	&	(e)	&	2	&	653.872	&	2.60E-06	&	2.29E+03	&	2.10E+03	&	-4.83	&	-4.87	&	0.053	\\
27602	&	(o)	&	3	&	6.1$^d$	&	68	&	(e)	&	2	&	363.074	&	7.66E-01	&	9.54E+08	&	8.79E+08	&	0.28	&	0.24	&	0.871	\\
	&		&		&		&	178	&	(e)	&	3	&	364.531	&	1.36E-01	&	1.69E+08	&	1.56E+08	&	-0.47	&	-0.51	&	0.952	\\
	&		&		&		&	2541	&	(e)	&	2	&	398.906	&	7.51E-04	&	9.35E+05	&	8.62E+05	&	-2.65	&	-2.69	&	0.768	\\
	&		&		&		&	4803	&	(e)	&	2	&	438.481	&	7.95E-03	&	9.90E+06	&	9.12E+06	&	-1.54	&	-1.58	&	0.966	\\
	&		&		&		&	4884	&	(e)	&	3	&	440.039	&	8.59E-02	&	1.07E+08	&	9.86E+07	&	-0.51	&	-0.54	&	0.975	\\
	&		&		&		&	4988	&	(e)	&	4	&	442.067	&	3.37E-03	&	4.20E+06	&	3.87E+06	&	-1.91	&	-1.95	&	0.242	\\
	&		&		&		&	10945	&	(e)	&	2	&	600.150	&	4.67E-06	&	5.82E+03	&	5.36E+03	&	-4.50	&	-4.54	&	0.520	\\
	&		&		&		&	12154	&	(e)	&	2	&	647.153	&	4.47E-05	&	5.57E+04	&	5.13E+04	&	-3.46	&	-3.49	&	0.612	\\
	&		&		&		&	14261	&	(e)	&	4	&	749.355	&	5.08E-08	&	6.33E+01	&	5.83E+01	&	-6.28	&	-6.31	&	0.089	\\
	27841	&	(o)	&	4	&	6.1$^d$	&	178	&	(e)	&	3	&	361.383	&	9.04E-01	&	1.47E+09	&	1.33E+09	&	0.46	&	0.42	&	0.949	\\
	&		&		&		&	4884	&	(e)	&	3	&	435.460	&	6.12E-03	&	9.96E+06	&	9.03E+06	&	-1.55	&	-1.59	&	0.976	\\
	&		&		&		&	4988	&	(e)	&	4	&	437.446	&	9.04E-02	&	1.47E+08	&	1.33E+08	&	-0.37	&	-0.42	&	0.976	\\
	&		&		&		&	14261	&	(e)	&	4	&	736.173	&	7.87E-07	&	1.28E+03	&	1.16E+03	&	-4.99	&	-5.03	&	0.912	\\
27918	&	(o)	&	1	&	4.7$^d$	&	0	&	(e)	&	1	&	358.092	&	5.67E-01	&	3.83E+08	&	3.62E+08	&	-0.13	&	-0.16	&	0.931	\\
	&		&		&		&	68	&	(e)	&	2	&	358.963	&	2.03E-01	&	1.37E+08	&	1.29E+08	&	-0.58	&	-0.60	&	0.950	\\
	&		&		&		&	2541	&	(e)	&	2	&	393.949	&	6.08E-06	&	4.11E+03	&	3.88E+03	&	-5.02	&	-5.04	&	0.001	\\
	&		&		&		&	4803	&	(e)	&	2	&	432.500	&	2.19E-01	&	1.48E+08	&	1.40E+08	&	-0.38	&	-0.41	&	0.961	\\
	&		&		&		&	10945	&	(e)	&	2	&	589.000	&	1.73E-04	&	1.17E+05	&	1.11E+05	&	-3.21	&	-3.24	&	0.214	\\
	&		&		&		&	11736	&	(e)	&	0	&	617.822	&	4.66E-04	&	3.15E+05	&	2.98E+05	&	-2.74	&	-2.77	&	0.704	\\
	&		&		&		&	12074	&	(e)	&	0	&	630.992	&	6.22E-03	&	4.20E+06	&	3.97E+06	&	-1.60	&	-1.63	&	0.611	\\
	&		&		&		&	12102	&	(e)	&	1	&	632.085	&	4.22E-03	&	2.85E+06	&	2.69E+06	&	-1.77	&	-1.79	&	0.563	\\
	&		&		&		&	12154	&	(e)	&	2	&	634.207	&	2.25E-04	&	1.52E+05	&	1.44E+05	&	-3.04	&	-3.06	&	0.323	\\
	&		&		&		&	25955	&	(e)	&	0	&	5093.945	&	6.48E-08	&	4.38E+01	&	4.14E+01	&	-4.72	&	-4.79	&	0.157	\\
28021	&	(o)	&	2	&	4.7$^d$	&	0	&	(e)	&	1	&	356.770	&	1.39E-01	&	1.57E+08	&	1.48E+08	&	-0.52	&	-0.55	&	0.950	\\
\hline
\end{tabular}
\end{small}
\end{center}
\label{gfsc2}
\end{table}
\end{landscape}

\newpage
\begin{landscape}
\addtocounter{table}{-1}
\begin{table}
\caption{Continued.}
\begin{center}
\begin{small}
\begin{tabular}{rrrrrrrrrrrrrr}
\hline
\multicolumn{3}{c}{Upper level$^a$} & \multicolumn{1}{c}{$\tau_u$ (ns)} & \multicolumn{3}{c}{Lower level$^a$}&
 \multicolumn{1}{c}{$\lambda^b$ (nm)} & \multicolumn{1}{c}{$BF$} & \multicolumn{1}{c}{$gA$ (s$^{-1}$)} & \multicolumn{1}{c}{$gA_{\text{resc}}$ (s$^{-1}$)} & \multicolumn{1}{c}{$\log(gf)$} & \multicolumn{1}{c}{$\log(gf)_{\text{resc}}$} & \multicolumn{1}{c}{CF$^c$} \\
\hline
	&		&		&		&	68	&	(e)	&	2	&	357.634	&	5.02E-01	&	5.68E+08	&	5.34E+08	&	0.04	&	0.01	&	0.857	\\
	&		&		&		&	178	&	(e)	&	3	&	359.047	&	1.27E-01	&	1.44E+08	&	1.35E+08	&	-0.55	&	-0.58	&	0.923	\\
	&		&		&		&	2541	&	(e)	&	2	&	392.348	&	2.16E-03	&	2.44E+06	&	2.29E+06	&	-2.25	&	-2.28	&	0.741	\\
	&		&		&		&	4803	&	(e)	&	2	&	430.571	&	2.13E-02	&	2.41E+07	&	2.26E+07	&	-1.17	&	-1.20	&	0.721	\\
	&		&		&		&	4884	&	(e)	&	3	&	432.073	&	1.98E-01	&	2.24E+08	&	2.10E+08	&	-0.20	&	-0.23	&	0.963	\\
	&		&		&		&	10945	&	(e)	&	2	&	585.430	&	8.92E-07	&	1.01E+03	&	9.49E+02	&	-5.28	&	-5.31	&	0.002	\\
	&		&		&		&	12102	&	(e)	&	1	&	627.975	&	8.83E-03	&	1.00E+07	&	9.40E+06	&	-1.23	&	-1.25	&	0.651	\\
	&		&		&		&	12154	&	(e)	&	2	&	630.070	&	2.24E-03	&	2.54E+06	&	2.39E+06	&	-1.82	&	-1.85	&	0.443	\\
28161	&	(o)	&	3	&	4.7$^d$	&	68	&	(e)	&	2	&	355.853	&	1.18E-01	&	1.88E+08	&	1.76E+08	&	-0.45	&	-0.48	&	0.968	\\
	&		&		&		&	178	&	(e)	&	3	&	357.253	&	6.52E-01	&	1.04E+09	&	9.72E+08	&	0.30	&	0.27	&	0.894	\\
	&		&		&		&	2541	&	(e)	&	2	&	390.206	&	2.31E-05	&	3.69E+04	&	3.45E+04	&	-4.08	&	-4.10	&	0.023	\\
	&		&		&		&	4803	&	(e)	&	2	&	427.993	&	3.06E-04	&	4.88E+05	&	4.56E+05	&	-2.87	&	-2.90	&	0.383	\\
	&		&		&		&	4884	&	(e)	&	3	&	429.477	&	1.37E-02	&	2.18E+07	&	2.04E+07	&	-1.22	&	-1.25	&	0.601	\\
	&		&		&		&	4988	&	(e)	&	4	&	431.408	&	2.04E-01	&	3.26E+08	&	3.05E+08	&	-0.04	&	-0.07	&	0.963	\\
	&		&		&		&	10945	&	(e)	&	2	&	580.673	&	1.57E-04	&	2.50E+05	&	2.34E+05	&	-2.90	&	-2.93	&	0.555	\\
	&		&		&		&	12154	&	(e)	&	2	&	624.564	&	1.10E-02	&	1.76E+07	&	1.64E+07	&	-0.99	&	-1.02	&	0.638	\\
	&		&		&		&	14261	&	(e)	&	4	&	719.234	&	3.06E-05	&	4.88E+04	&	4.56E+04	&	-3.42	&	-3.45	&	0.419	\\
29736	&	(o)	&	0	&	7.7$^d$	&	0	&	(e)	&	1	&	336.193	&	8.75E-01	&	1.38E+08	&	1.14E+08	&	-0.63	&	-0.72	&	0.519	\\
	&		&		&		&	12102	&	(e)	&	1	&	566.904	&	1.25E-01	&	1.98E+07	&	1.63E+07	&	-1.02	&	-1.10	&	0.885	\\
29742	&	(o)	&	1	&	7.6$^d$	&	0	&	(e)	&	1	&	336.127	&	2.43E-01	&	1.14E+08	&	9.57E+07	&	-0.72	&	-0.79	&	0.538	\\
	&		&		&		&	68	&	(e)	&	2	&	336.894	&	6.21E-01	&	2.92E+08	&	2.45E+08	&	-0.30	&	-0.38	&	0.485	\\
	&		&		&		&	2541	&	(e)	&	2	&	367.526	&	4.00E-03	&	1.88E+06	&	1.58E+06	&	-2.42	&	-2.49	&	0.070	\\
	&		&		&		&	4803	&	(e)	&	2	&	400.860	&	2.21E-04	&	1.04E+05	&	8.73E+04	&	-3.60	&	-3.68	&	0.479	\\
	&		&		&		&	10945	&	(e)	&	2	&	531.835	&	8.60E-03	&	4.04E+06	&	3.39E+06	&	-1.77	&	-1.84	&	0.723	\\
	&		&		&		&	11736	&	(e)	&	0	&	555.222	&	5.30E-03	&	2.49E+06	&	2.09E+06	&	-1.94	&	-2.01	&	0.771	\\
	&		&		&		&	12074	&	(e)	&	0	&	565.836	&	3.72E-02	&	1.75E+07	&	1.47E+07	&	-1.08	&	-1.15	&	0.736	\\
	&		&		&		&	12102	&	(e)	&	1	&	566.715	&	3.19E-02	&	1.50E+07	&	1.26E+07	&	-1.15	&	-1.22	&	0.881	\\
	&		&		&		&	12154	&	(e)	&	2	&	568.420	&	4.89E-02	&	2.30E+07	&	1.93E+07	&	-0.96	&	-1.03	&	0.813	\\
	&		&		&		&	25955	&	(e)	&	0	&	2639.920	&	7.02E-06	&	3.30E+03	&	2.77E+03	&	-3.46	&	-3.54	&	0.175	\\
29824	&	(o)	&	2	&	7.4$^d$	&	0	&	(e)	&	1	&	335.205	&	1.05E-02	&	8.36E+06	&	7.12E+06	&	-1.85	&	-1.92	&	0.529	\\
	&		&		&		&	68	&	(e)	&	2	&	335.968	&	1.47E-01	&	1.17E+08	&	9.96E+07	&	-0.71	&	-0.77	&	0.537	\\
	&		&		&		&	178	&	(e)	&	3	&	337.215	&	7.13E-01	&	5.66E+08	&	4.82E+08	&	-0.02	&	-0.09	&	0.506	\\
	&		&		&		&	2541	&	(e)	&	2	&	366.425	&	2.42E-03	&	1.92E+06	&	1.63E+06	&	-2.42	&	-2.48	&	0.731	\\
	&		&		&		&	4803	&	(e)	&	2	&	399.550	&	3.24E-05	&	2.57E+04	&	2.19E+04	&	-4.21	&	-4.28	&	0.865	\\
	&		&		&		&	4884	&	(e)	&	3	&	400.843	&	1.69E-04	&	1.34E+05	&	1.14E+05	&	-3.49	&	-3.56	&	0.912	\\
	&		&		&		&	10945	&	(e)	&	2	&	529.531	&	4.02E-04	&	3.19E+05	&	2.72E+05	&	-2.88	&	-2.94	&	0.210	\\
	&		&		&		&	12102	&	(e)	&	1	&	564.100	&	3.08E-02	&	2.44E+07	&	2.08E+07	&	-0.94	&	-1.00	&	0.825	\\
	&		&		&		&	12154	&	(e)	&	2	&	565.790	&	9.49E-02	&	7.53E+07	&	6.41E+07	&	-0.45	&	-0.51	&	0.879	\\
30816	&	(o)	&	1	&	8.8$^d$	&	0	&	(e)	&	1	&	324.416	&	7.98E-04	&	2.95E+05	&	2.72E+05	&	-3.33	&	-3.37	&	0.025	\\
	&		&		&		&	68	&	(e)	&	2	&	325.131	&	2.31E-02	&	8.56E+06	&	7.89E+06	&	-1.87	&	-1.90	&	0.259	\\
	&		&		&		&	2541	&	(e)	&	2	&	353.571	&	4.73E-01	&	1.75E+08	&	1.61E+08	&	-0.49	&	-0.52	&	0.193	\\
	&		&		&		&	4803	&	(e)	&	2	&	384.317	&	2.68E-04	&	9.91E+04	&	9.14E+04	&	-3.66	&	-3.69	&	0.065	\\
	&		&		&		&	10945	&	(e)	&	2	&	503.102	&	3.62E-01	&	1.34E+08	&	1.24E+08	&	-0.29	&	-0.33	&	0.723	\\
	&		&		&		&	11736	&	(e)	&	0	&	523.981	&	1.24E-01	&	4.60E+07	&	4.24E+07	&	-0.72	&	-0.76	&	0.685	\\
\hline
\end{tabular}
\end{small}
\end{center}
\label{gfsc2}
\end{table}
\end{landscape}

\newpage
\begin{landscape}
\addtocounter{table}{-1}
\begin{table}
\caption{Continued.}
\begin{center}
\begin{small}
\begin{tabular}{rrrrrrrrrrrrrr}
\hline
\multicolumn{3}{c}{Upper level$^a$} & \multicolumn{1}{c}{$\tau_u$ (ns)} & \multicolumn{3}{c}{Lower level$^a$}&
 \multicolumn{1}{c}{$\lambda^b$ (nm)} & \multicolumn{1}{c}{$BF$} & \multicolumn{1}{c}{$gA$ (s$^{-1}$)} & \multicolumn{1}{c}{$gA_{\text{resc}}$ (s$^{-1}$)} & \multicolumn{1}{c}{$\log(gf)$} & \multicolumn{1}{c}{$\log(gf)_{\text{resc}}$} & \multicolumn{1}{c}{CF$^c$} \\
\hline
	&		&		&		&	12074	&	(e)	&	0	&	533.424	&	8.14E-03	&	3.01E+06	&	2.77E+06	&	-1.89	&	-1.93	&	0.773	\\
	&		&		&		&	12102	&	(e)	&	1	&	534.205	&	7.73E-04	&	2.86E+05	&	2.64E+05	&	-2.91	&	-2.95	&	0.422	\\
	&		&		&		&	12154	&	(e)	&	2	&	535.720	&	6.06E-03	&	2.24E+06	&	2.06E+06	&	-2.01	&	-2.05	&	0.758	\\
	&		&		&		&	25955	&	(e)	&	0	&	2056.840	&	8.71E-04	&	3.22E+05	&	2.97E+05	&	-1.67	&	-1.73	&	0.214	\\
32350	&	(o)	&	3	&	5.1$^d$	&	68	&	(e)	&	2	&	309.678	&	5.40E-04	&	8.08E+05	&	7.41E+05	&	-2.94	&	-2.97	&	0.182	\\
	&		&		&		&	178	&	(e)	&	3	&	310.737	&	5.99E-04	&	8.97E+05	&	8.22E+05	&	-2.89	&	-2.92	&	0.835	\\
	&		&		&		&	2541	&	(e)	&	2	&	335.372	&	7.22E-01	&	1.08E+09	&	9.90E+08	&	0.26	&	0.22	&	0.640	\\
	&		&		&		&	4803	&	(e)	&	2	&	362.911	&	2.65E-04	&	3.96E+05	&	3.63E+05	&	-3.11	&	-3.14	&	0.722	\\
	&		&		&		&	4884	&	(e)	&	3	&	363.977	&	1.13E-05	&	1.69E+04	&	1.55E+04	&	-4.47	&	-4.51	&	0.411	\\
	&		&		&		&	4988	&	(e)	&	4	&	365.364	&	9.09E-05	&	1.36E+05	&	1.25E+05	&	-3.57	&	-3.60	&	0.222	\\
	&		&		&		&	10945	&	(e)	&	2	&	467.041	&	8.42E-02	&	1.26E+08	&	1.16E+08	&	-0.39	&	-0.42	&	0.625	\\
	&		&		&		&	12154	&	(e)	&	2	&	495.020	&	4.18E-04	&	6.25E+05	&	5.73E+05	&	-2.64	&	-2.68	&	0.621	\\
	&		&		&		&	14261	&	(e)	&	4	&	552.679	&	1.92E-01	&	2.88E+08	&	2.64E+08	&	0.12	&	0.08	&	0.917	\\
39002	&	(o)	&	0	&	3.7$^d$	&	0	&	(e)	&	1	&	256.319	&	9.94E-01	&	2.70E+08	&	2.69E+08	&	-0.58	&	-0.58	&	0.958	\\
	&		&		&		&	12102	&	(e)	&	1	&	371.632	&	5.85E-03	&	1.59E+06	&	1.58E+06	&	-2.48	&	-2.48	&	0.139	\\
39115	&	(o)	&	1	&	3.7$^d$	&	0	&	(e)	&	1	&	255.580	&	2.50E-01	&	2.04E+08	&	2.03E+08	&	-0.70	&	-0.70	&	0.958	\\
	&		&		&		&	68	&	(e)	&	2	&	256.023	&	7.41E-01	&	6.04E+08	&	6.01E+08	&	-0.23	&	-0.23	&	0.955	\\
	&		&		&		&	2541	&	(e)	&	2	&	273.337	&	4.16E-04	&	3.39E+05	&	3.37E+05	&	-3.42	&	-3.42	&	0.244	\\
	&		&		&		&	4803	&	(e)	&	2	&	291.357	&	4.23E-07	&	3.45E+02	&	3.43E+02	&	-6.36	&	-6.36	&	0.362	\\
	&		&		&		&	10945	&	(e)	&	2	&	354.880	&	2.80E-05	&	2.28E+04	&	2.27E+04	&	-4.37	&	-4.37	&	0.095	\\
	&		&		&		&	11736	&	(e)	&	0	&	365.144	&	5.13E-07	&	4.18E+02	&	4.16E+02	&	-6.08	&	-6.08	&	0.001	\\
	&		&		&		&	12074	&	(e)	&	0	&	369.704	&	1.94E-03	&	1.58E+06	&	1.57E+06	&	-2.49	&	-2.49	&	0.137	\\
	&		&		&		&	12102	&	(e)	&	1	&	370.079	&	1.44E-03	&	1.17E+06	&	1.16E+06	&	-2.62	&	-2.62	&	0.137	\\
	&		&		&		&	12154	&	(e)	&	2	&	370.806	&	2.36E-03	&	1.92E+06	&	1.91E+06	&	-2.40	&	-2.40	&	0.136	\\
	&		&		&		&	25955	&	(e)	&	0	&	759.679	&	5.27E-07	&	4.29E+02	&	4.27E+02	&	-5.43	&	-5.43	&	0.074	\\
39346	&	(o)	&	2	&	3.8$^d$	&	0	&	(e)	&	1	&	254.082	&	1.02E-02	&	1.38E+07	&	1.34E+07	&	-1.88	&	-1.89	&	0.957	\\
	&		&		&		&	68	&	(e)	&	2	&	254.520	&	1.50E-01	&	2.04E+08	&	1.98E+08	&	-0.70	&	-0.72	&	0.957	\\
	&		&		&		&	178	&	(e)	&	3	&	255.235	&	8.34E-01	&	1.13E+09	&	1.10E+09	&	0.04	&	0.03	&	0.956	\\
	&		&		&		&	2541	&	(e)	&	2	&	271.625	&	2.29E-04	&	3.10E+05	&	3.01E+05	&	-3.47	&	-3.48	&	0.829	\\
	&		&		&		&	4803	&	(e)	&	2	&	289.412	&	1.49E-07	&	2.02E+02	&	1.96E+02	&	-6.60	&	-6.61	&	0.745	\\
	&		&		&		&	4884	&	(e)	&	3	&	290.090	&	4.46E-07	&	6.05E+02	&	5.87E+02	&	-6.12	&	-6.13	&	0.917	\\
	&		&		&		&	10945	&	(e)	&	2	&	352.000	&	2.32E-05	&	3.14E+04	&	3.05E+04	&	-4.23	&	-4.25	&	0.097	\\
	&		&		&		&	12102	&	(e)	&	1	&	366.949	&	1.36E-03	&	1.85E+06	&	1.80E+06	&	-2.43	&	-2.44	&	0.129	\\
	&		&		&		&	12154	&	(e)	&	2	&	367.663	&	4.12E-03	&	5.58E+06	&	5.42E+06	&	-1.95	&	-1.96	&	0.132	\\
57744	&	(e)	&	3	&	3.20$^e$	&	26081	&	(o)	&	2	&	315.739	&	5.95E-06	&	1.19E+04	&	1.30E+04	&	-4.75	&	-4.71	&	0.004	\\
	&		&		&		&	27444	&	(o)	&	2	&	329.936	&	1.75E-03	&	3.49E+06	&	3.82E+06	&	-2.25	&	-2.21	&	0.660	\\
	&		&		&		&	27602	&	(o)	&	3	&	331.673	&	4.87E-02	&	9.74E+07	&	1.07E+08	&	-0.80	&	-0.75	&	0.694	\\
	&		&		&		&	27841	&	(o)	&	4	&	334.323	&	3.77E-01	&	7.53E+08	&	8.24E+08	&	0.10	&	0.14	&	0.668	\\
	&		&		&		&	28021	&	(o)	&	2	&	336.347	&	5.55E-02	&	1.11E+08	&	1.21E+08	&	-0.73	&	-0.69	&	0.851	\\
	&		&		&		&	28161	&	(o)	&	3	&	337.938	&	3.93E-01	&	7.86E+08	&	8.60E+08	&	0.13	&	0.17	&	0.834	\\
	&		&		&		&	29824	&	(o)	&	2	&	358.064	&	1.24E-01	&	2.47E+08	&	2.70E+08	&	-0.32	&	-0.28	&	0.458	\\
	&		&		&		&	32350	&	(o)	&	3	&	393.683	&	8.95E-05	&	1.79E+05	&	1.96E+05	&	-3.38	&	-3.34	&	0.357	\\
	&		&		&		&	39346	&	(o)	&	2	&	543.375	&	5.65E-04	&	1.13E+06	&	1.24E+06	&	-2.30	&	-2.26	&	0.027	\\
58252	&	(e)	&	2	&	3.26$^e$	&	26081	&	(o)	&	2	&	310.751	&	3.70E-01	&	4.99E+08	&	5.67E+08	&	-0.14	&	-0.09	&	0.549	\\
\hline
\end{tabular}
\end{small}
\end{center}
\label{gfsc2}
\end{table}
\end{landscape}

\newpage
\begin{landscape}
\addtocounter{table}{-1}
\begin{table}
\caption{Continued.}
\begin{center}
\begin{small}
\begin{tabular}{rrrrrrrrrrrrrr}
\hline
\multicolumn{3}{c}{Upper level$^a$} & \multicolumn{1}{c}{$\tau_u$ (ns)} & \multicolumn{3}{c}{Lower level$^a$}&
 \multicolumn{1}{c}{$\lambda^b$ (nm)} & \multicolumn{1}{c}{$BF$} & \multicolumn{1}{c}{$gA$ (s$^{-1}$)} & \multicolumn{1}{c}{$gA_{\text{resc}}$ (s$^{-1}$)} & \multicolumn{1}{c}{$\log(gf)$} & \multicolumn{1}{c}{$\log(gf)_{\text{resc}}$} & \multicolumn{1}{c}{CF$^c$} \\
\hline
	&		&		&		&	27444	&	(o)	&	2	&	324.493	&	9.48E-03	&	1.28E+07	&	1.45E+07	&	-1.70	&	-1.64	&	0.583	\\
	&		&		&		&	27602	&	(o)	&	3	&	326.174	&	7.63E-03	&	1.03E+07	&	1.17E+07	&	-1.79	&	-1.73	&	0.542	\\
	&		&		&		&	27918	&	(o)	&	1	&	329.565	&	3.92E-04	&	5.29E+05	&	6.01E+05	&	-3.07	&	-3.01	&	0.062	\\
	&		&		&		&	28021	&	(o)	&	2	&	330.693	&	1.11E-02	&	1.50E+07	&	1.70E+07	&	-1.61	&	-1.55	&	0.754	\\
	&		&		&		&	28161	&	(o)	&	3	&	332.231	&	7.48E-04	&	1.01E+06	&	1.15E+06	&	-2.78	&	-2.72	&	0.136	\\
	&		&		&		&	29742	&	(o)	&	1	&	350.655	&	4.07E-04	&	5.50E+05	&	6.25E+05	&	-2.99	&	-2.94	&	0.016	\\
	&		&		&		&	29824	&	(o)	&	2	&	351.663	&	1.91E-03	&	2.58E+06	&	2.93E+06	&	-2.32	&	-2.26	&	0.439	\\
	&		&		&		&	30816	&	(o)	&	1	&	364.376	&	1.56E-01	&	2.10E+08	&	2.39E+08	&	-0.38	&	-0.32	&	0.524	\\
	&		&		&		&	32350	&	(o)	&	3	&	385.960	&	4.43E-01	&	5.98E+08	&	6.79E+08	&	0.13	&	0.18	&	0.811	\\
	&		&		&		&	39115	&	(o)	&	1	&	522.401	&	6.04E-07	&	8.16E+02	&	9.27E+02	&	-5.48	&	-5.42	&	0.001	\\
	&		&		&		&	39346	&	(o)	&	2	&	528.770	&	5.48E-06	&	7.40E+03	&	8.41E+03	&	-4.51	&	-4.45	&	0.032	\\
	&		&		&		&	55715	&	(o)	&	1	&	3941.008	&	1.07E-04	&	1.45E+05	&	1.65E+05	&	-1.48	&	-1.42	&	0.272	\\
59528	&	(e)	&	3	&	2.32	$^e$&	26081	&	(o)	&	2	&	298.893	&	7.30E-01	&	1.90E+09	&	2.20E+09	&	0.41	&	0.47	&	0.813	\\
	&		&		&		&	27444	&	(o)	&	2	&	311.585	&	5.11E-04	&	1.33E+06	&	1.54E+06	&	-2.71	&	-2.65	&	0.017	\\
	&		&		&		&	27602	&	(o)	&	3	&	313.134	&	1.28E-03	&	3.34E+06	&	3.87E+06	&	-2.31	&	-2.24	&	0.483	\\
	&		&		&		&	27841	&	(o)	&	4	&	315.495	&	1.30E-03	&	3.38E+06	&	3.92E+06	&	-2.30	&	-2.23	&	0.444	\\
	&		&		&		&	28021	&	(o)	&	2	&	317.297	&	8.37E-05	&	2.18E+05	&	2.53E+05	&	-3.48	&	-3.42	&	0.013	\\
	&		&		&		&	28161	&	(o)	&	3	&	318.712	&	7.76E-03	&	2.02E+07	&	2.34E+07	&	-1.51	&	-1.45	&	0.493	\\
	&		&		&		&	29824	&	(o)	&	2	&	336.553	&	2.29E-03	&	5.97E+06	&	6.92E+06	&	-1.99	&	-1.93	&	0.229	\\
	&		&		&		&	32350	&	(o)	&	3	&	367.834	&	2.57E-01	&	6.69E+08	&	7.75E+08	&	0.14	&	0.20	&	0.867	\\
	&		&		&		&	39346	&	(o)	&	2	&	495.331	&	1.85E-04	&	4.81E+05	&	5.57E+05	&	-2.75	&	-2.69	&	0.621	\\
59875	&	(e)	&	1	&	2.23$^e$	&	26081	&	(o)	&	2	&	295.826	&	1.55E-02	&	1.71E+07	&	2.09E+07	&	-1.64	&	-1.56	&	0.658	\\
	&		&		&		&	27444	&	(o)	&	2	&	308.254	&	1.22E-01	&	1.34E+08	&	1.64E+08	&	-0.72	&	-0.63	&	0.631	\\
	&		&		&		&	27918	&	(o)	&	1	&	312.827	&	3.93E-01	&	4.33E+08	&	5.29E+08	&	-0.20	&	-0.11	&	0.597	\\
	&		&		&		&	28021	&	(o)	&	2	&	313.843	&	1.12E-01	&	1.23E+08	&	1.50E+08	&	-0.74	&	-0.65	&	0.481	\\
	&		&		&		&	29736	&	(o)	&	0	&	331.703	&	1.92E-01	&	2.11E+08	&	2.58E+08	&	-0.46	&	-0.37	&	0.899	\\
	&		&		&		&	29742	&	(o)	&	1	&	331.768	&	1.42E-01	&	1.57E+08	&	1.92E+08	&	-0.59	&	-0.50	&	0.750	\\
	&		&		&		&	29824	&	(o)	&	2	&	332.670	&	7.03E-03	&	7.74E+06	&	9.45E+06	&	-1.89	&	-1.80	&	0.463	\\
	&		&		&		&	30816	&	(o)	&	1	&	344.024	&	9.17E-04	&	1.01E+06	&	1.23E+06	&	-2.75	&	-2.66	&	0.021	\\
	&		&		&		&	39002	&	(o)	&	0	&	478.957	&	9.08E-03	&	1.00E+07	&	1.22E+07	&	-1.46	&	-1.38	&	0.769	\\
	&		&		&		&	39115	&	(o)	&	1	&	481.560	&	6.81E-03	&	7.50E+06	&	9.16E+06	&	-1.58	&	-1.50	&	0.766	\\
	&		&		&		&	39346	&	(o)	&	2	&	486.967	&	3.60E-04	&	3.97E+05	&	4.85E+05	&	-2.85	&	-2.76	&	0.542	\\
	&		&		&		&	55715	&	(o)	&	1	&	2403.352	&	2.00E-05	&	2.20E+04	&	2.69E+04	&	-2.72	&	-2.63	&	0.500	\\
59929	&	(e)	&	2	&	2.32	$^e$&	26081	&	(o)	&	2	&	295.351	&	5.27E-05	&	9.68E+04	&	1.14E+05	&	-3.89	&	-3.83	&	0.011	\\
	&		&		&		&	27444	&	(o)	&	2	&	307.738	&	9.53E-03	&	1.75E+07	&	2.05E+07	&	-1.61	&	-1.53	&	0.294	\\
	&		&		&		&	27602	&	(o)	&	3	&	309.249	&	1.16E-01	&	2.13E+08	&	2.50E+08	&	-0.52	&	-0.45	&	0.667	\\
	&		&		&		&	27918	&	(o)	&	1	&	312.296	&	7.46E-02	&	1.37E+08	&	1.61E+08	&	-0.70	&	-0.63	&	0.512	\\
	&		&		&		&	28021	&	(o)	&	2	&	313.309	&	3.62E-01	&	6.64E+08	&	7.79E+08	&	-0.01	&	0.06	&	0.578	\\
	&		&		&		&	28161	&	(o)	&	3	&	314.688	&	6.48E-02	&	1.19E+08	&	1.40E+08	&	-0.76	&	-0.68	&	0.420	\\
	&		&		&		&	29742	&	(o)	&	1	&	331.170	&	2.71E-01	&	4.97E+08	&	5.83E+08	&	-0.09	&	-0.02	&	0.907	\\
	&		&		&		&	29824	&	(o)	&	2	&	332.069	&	7.73E-02	&	1.42E+08	&	1.67E+08	&	-0.63	&	-0.56	&	0.690	\\
	&		&		&		&	30816	&	(o)	&	1	&	343.382	&	3.84E-03	&	7.06E+06	&	8.29E+06	&	-1.91	&	-1.83	&	0.362	\\
	&		&		&		&	32350	&	(o)	&	3	&	362.485	&	4.89E-06	&	8.98E+03	&	1.05E+04	&	-4.75	&	-4.68	&	0.022	\\
	&		&		&		&	39115	&	(o)	&	1	&	480.302	&	1.24E-02	&	2.28E+07	&	2.68E+07	&	-1.10	&	-1.03	&	0.783	\\
\hline
\end{tabular}
\end{small}
\end{center}
\label{gfsc2}
\end{table}
\end{landscape}

\newpage
\begin{landscape}
\addtocounter{table}{-1}
\begin{table}
\caption{Continued.}
\begin{center}
\begin{small}
\begin{tabular}{rrrrrrrrrrrrrr}
\hline
\multicolumn{3}{c}{Upper level$^a$} & \multicolumn{1}{c}{$\tau_u$ (ns)} & \multicolumn{3}{c}{Lower level$^a$}&
 \multicolumn{1}{c}{$\lambda^b$ (nm)} & \multicolumn{1}{c}{$BF$} & \multicolumn{1}{c}{$gA$ (s$^{-1}$)} & \multicolumn{1}{c}{$gA_{\text{resc}}$ (s$^{-1}$)} & \multicolumn{1}{c}{$\log(gf)$} & \multicolumn{1}{c}{$\log(gf)_{\text{resc}}$} & \multicolumn{1}{c}{CF$^c$} \\
\hline
	&		&		&		&	39346	&	(o)	&	2	&	485.680	&	3.79E-03	&	6.96E+06	&	8.17E+06	&	-1.61	&	-1.54	&	0.716	\\
	&		&		&		&	55715	&	(o)	&	1	&	2372.344	&	1.32E-09	&	2.42E+00	&	2.84E+00	&	-6.69	&	-6.62	&	0.003	\\
60002	&	(e)	&	3	&	2.41$^e$	&	26081	&	(o)	&	2	&	294.720	&	5.91E-03	&	1.50E+07	&	1.72E+07	&	-1.71	&	-1.65	&	0.241	\\
	&		&		&		&	27444	&	(o)	&	2	&	307.053	&	9.57E-04	&	2.43E+06	&	2.78E+06	&	-2.47	&	-2.41	&	0.144	\\
	&		&		&		&	27602	&	(o)	&	3	&	308.558	&	3.10E-03	&	7.86E+06	&	9.00E+06	&	-1.95	&	-1.89	&	0.088	\\
	&		&		&		&	27841	&	(o)	&	4	&	310.850	&	1.09E-01	&	2.76E+08	&	3.16E+08	&	-0.40	&	-0.34	&	0.677	\\
	&		&		&		&	28021	&	(o)	&	2	&	312.599	&	5.52E-02	&	1.40E+08	&	1.60E+08	&	-0.69	&	-0.63	&	0.476	\\
	&		&		&		&	28161	&	(o)	&	3	&	313.972	&	4.41E-01	&	1.12E+09	&	1.28E+09	&	0.22	&	0.28	&	0.569	\\
	&		&		&		&	29824	&	(o)	&	2	&	331.272	&	3.63E-01	&	9.22E+08	&	1.06E+09	&	0.18	&	0.24	&	0.916	\\
	&		&		&		&	32350	&	(o)	&	3	&	361.535	&	5.75E-03	&	1.46E+07	&	1.67E+07	&	-1.54	&	-1.48	&	0.807	\\
	&		&		&		&	39346	&	(o)	&	2	&	483.977	&	1.58E-02	&	4.00E+07	&	4.58E+07	&	-0.85	&	-0.79	&	0.794	\\
60267	&	(e)	&	3	&	2.19	$^e$&	26081	&	(o)	&	2	&	292.433	&	2.06E-03	&	5.78E+06	&	6.59E+06	&	-2.13	&	-2.07	&	0.062	\\
	&		&		&		&	27444	&	(o)	&	2	&	304.572	&	9.17E-01	&	2.57E+09	&	2.93E+09	&	0.55	&	0.61	&	0.862	\\
	&		&		&		&	27602	&	(o)	&	3	&	306.052	&	7.60E-02	&	2.13E+08	&	2.43E+08	&	-0.52	&	-0.47	&	0.784	\\
	&		&		&		&	27841	&	(o)	&	4	&	308.307	&	7.21E-04	&	2.02E+06	&	2.30E+06	&	-2.54	&	-2.48	&	0.321	\\
	&		&		&		&	28021	&	(o)	&	2	&	310.027	&	1.66E-03	&	4.65E+06	&	5.30E+06	&	-2.17	&	-2.12	&	0.469	\\
	&		&		&		&	28161	&	(o)	&	3	&	311.378	&	9.67E-04	&	2.71E+06	&	3.09E+06	&	-2.40	&	-2.35	&	0.618	\\
	&		&		&		&	29824	&	(o)	&	2	&	328.386	&	2.14E-04	&	5.99E+05	&	6.83E+05	&	-3.01	&	-2.96	&	0.858	\\
	&		&		&		&	32350	&	(o)	&	3	&	358.100	&	1.44E-03	&	4.03E+06	&	4.60E+06	&	-2.11	&	-2.05	&	0.870	\\
	&		&		&		&	39346	&	(o)	&	2	&	477.840	&	6.67E-06	&	1.87E+04	&	2.13E+04	&	-4.19	&	-4.14	&	0.781	\\
60348	&	(e)	&	4	&	2.4$^d$	&	27602	&	(o)	&	3	&	305.292	&	9.37E-01	&	3.35E+09	&	3.52E+09	&	0.67	&	0.69	&	0.863	\\
	&		&		&		&	27841	&	(o)	&	4	&	307.536	&	5.82E-02	&	2.08E+08	&	2.18E+08	&	-0.53	&	-0.51	&	0.781	\\
	&		&		&		&	28161	&	(o)	&	3	&	310.592	&	4.25E-03	&	1.52E+07	&	1.60E+07	&	-1.66	&	-1.64	&	0.846	\\
	&		&		&		&	32350	&	(o)	&	3	&	357.060	&	7.98E-05	&	2.85E+05	&	2.99E+05	&	-3.26	&	-3.24	&	0.295	\\
60400	&	(e)	&	1	&	2.44$^e$	&	26081	&	(o)	&	2	&	291.298	&	2.97E-01	&	3.08E+08	&	3.65E+08	&	-0.41	&	-0.33	&	0.746	\\
	&		&		&		&	27444	&	(o)	&	2	&	303.340	&	1.18E-02	&	1.23E+07	&	1.46E+07	&	-1.77	&	-1.70	&	0.731	\\
	&		&		&		&	27918	&	(o)	&	1	&	307.767	&	2.40E-03	&	2.49E+06	&	2.95E+06	&	-2.45	&	-2.38	&	0.076	\\
	&		&		&		&	28021	&	(o)	&	2	&	308.751	&	4.66E-03	&	4.84E+06	&	5.73E+06	&	-2.16	&	-2.09	&	0.499	\\
	&		&		&		&	29736	&	(o)	&	0	&	326.020	&	1.01E-02	&	1.05E+07	&	1.24E+07	&	-1.78	&	-1.70	&	0.891	\\
	&		&		&		&	29742	&	(o)	&	1	&	326.082	&	1.70E-02	&	1.76E+07	&	2.08E+07	&	-1.55	&	-1.48	&	0.255	\\
	&		&		&		&	29824	&	(o)	&	2	&	326.955	&	3.85E-03	&	4.00E+06	&	4.74E+06	&	-2.19	&	-2.12	&	0.314	\\
	&		&		&		&	30816	&	(o)	&	1	&	337.915	&	6.51E-01	&	6.76E+08	&	8.01E+08	&	0.06	&	0.14	&	0.880	\\
	&		&		&		&	39002	&	(o)	&	0	&	467.198	&	5.28E-04	&	5.48E+05	&	6.49E+05	&	-2.75	&	-2.67	&	0.735	\\
	&		&		&		&	39115	&	(o)	&	1	&	469.675	&	1.63E-05	&	1.69E+04	&	2.00E+04	&	-4.26	&	-4.18	&	0.014	\\
	&		&		&		&	39346	&	(o)	&	2	&	474.816	&	3.58E-04	&	3.72E+05	&	4.41E+05	&	-2.90	&	-2.83	&	0.590	\\
	&		&		&		&	55715	&	(o)	&	1	&	2133.866	&	1.43E-03	&	1.48E+06	&	1.75E+06	&	-1.00	&	-0.92	&	0.554	\\
60457	&	(e)	&	5	&	2.5$^d$	&	27841	&	(o)	&	4	&	306.511	&	1.00E+00	&	4.33E+09	&	4.40E+09	&	0.79	&	0.79	&	0.865	\\
61071	&	(e)	&	1	&	2.45	$^e$&	26081	&	(o)	&	2	&	285.711	&	6.87E-04	&	7.30E+05	&	8.41E+05	&	-3.05	&	-2.99	&	0.084	\\
	&		&		&		&	27444	&	(o)	&	2	&	297.287	&	1.30E-05	&	1.38E+04	&	1.59E+04	&	-4.74	&	-4.68	&	0.061	\\
	&		&		&		&	27918	&	(o)	&	1	&	301.538	&	8.24E-05	&	8.76E+04	&	1.01E+05	&	-3.93	&	-3.86	&	0.091	\\
	&		&		&		&	28021	&	(o)	&	2	&	302.483	&	6.95E-05	&	7.39E+04	&	8.51E+04	&	-4.00	&	-3.93	&	0.056	\\
	&		&		&		&	29736	&	(o)	&	0	&	319.038	&	1.06E-01	&	1.13E+08	&	1.30E+08	&	-0.76	&	-0.70	&	0.841	\\
	&		&		&		&	29742	&	(o)	&	1	&	319.098	&	3.00E-01	&	3.19E+08	&	3.68E+08	&	-0.31	&	-0.25	&	0.802	\\
	&		&		&		&	29824	&	(o)	&	2	&	319.933	&	5.06E-01	&	5.38E+08	&	6.20E+08	&	-0.08	&	-0.02	&	0.837	\\
\hline
\end{tabular}
\end{small}
\end{center}
\label{gfsc2}
\end{table}
\end{landscape}

\newpage
\begin{landscape}
\addtocounter{table}{-1}
\begin{table}
\caption{Continued.}
\begin{center}
\begin{small}
\begin{tabular}{rrrrrrrrrrrrrr}
\hline
\multicolumn{3}{c}{Upper level$^a$} & \multicolumn{1}{c}{$\tau_u$ (ns)} & \multicolumn{3}{c}{Lower level$^a$}&
 \multicolumn{1}{c}{$\lambda^b$ (nm)} & \multicolumn{1}{c}{$BF$} & \multicolumn{1}{c}{$gA$ (s$^{-1}$)} & \multicolumn{1}{c}{$gA_{\text{resc}}$ (s$^{-1}$)} & \multicolumn{1}{c}{$\log(gf)$} & \multicolumn{1}{c}{$\log(gf)_{\text{resc}}$} & \multicolumn{1}{c}{CF$^c$} \\
\hline
	&		&		&		&	30816	&	(o)	&	1	&	330.421	&	2.64E-02	&	2.81E+07	&	3.24E+07	&	-1.34	&	-1.28	&	0.895	\\
	&		&		&		&	39002	&	(o)	&	0	&	452.993	&	6.99E-03	&	7.43E+06	&	8.56E+06	&	-1.64	&	-1.58	&	0.718	\\
	&		&		&		&	39115	&	(o)	&	1	&	455.321	&	2.08E-02	&	2.21E+07	&	2.55E+07	&	-1.17	&	-1.10	&	0.750	\\
	&		&		&		&	39346	&	(o)	&	2	&	460.151	&	3.23E-02	&	3.43E+07	&	3.95E+07	&	-0.96	&	-0.90	&	0.742	\\
	&		&		&		&	55715	&	(o)	&	1	&	1866.531	&	1.98E-05	&	2.10E+04	&	2.42E+04	&	-2.96	&	-2.90	&	0.528	\\
63375	&	(e)	&	2	&	2.15	$^e$&	26081	&	(o)	&	2	&	268.065	&	1.38E-02	&	2.88E+07	&	3.21E+07	&	-1.51	&	-1.46	&	0.561	\\
	&		&		&		&	27444	&	(o)	&	2	&	278.230	&	3.17E-01	&	6.61E+08	&	7.37E+08	&	-0.12	&	-0.07	&	0.673	\\
	&		&		&		&	27602	&	(o)	&	3	&	279.464	&	4.12E-02	&	8.59E+07	&	9.58E+07	&	-1.00	&	-0.95	&	0.701	\\
	&		&		&		&	27918	&	(o)	&	1	&	281.950	&	5.37E-01	&	1.12E+09	&	1.25E+09	&	0.12	&	0.17	&	0.685	\\
	&		&		&		&	28021	&	(o)	&	2	&	282.776	&	8.92E-02	&	1.86E+08	&	2.07E+08	&	-0.65	&	-0.60	&	0.516	\\
	&		&		&		&	28161	&	(o)	&	3	&	283.899	&	1.87E-03	&	3.91E+06	&	4.36E+06	&	-2.33	&	-2.28	&	0.263	\\
	&		&		&		&	29742	&	(o)	&	1	&	297.245	&	3.20E-05	&	6.68E+04	&	7.45E+04	&	-4.05	&	-4.01	&	0.023	\\
	&		&		&		&	29824	&	(o)	&	2	&	297.969	&	2.25E-05	&	4.70E+04	&	5.24E+04	&	-4.20	&	-4.16	&	0.075	\\
	&		&		&		&	30816	&	(o)	&	1	&	307.046	&	2.30E-06	&	4.79E+03	&	5.34E+03	&	-5.17	&	-5.12	&	0.000	\\
	&		&		&		&	32350	&	(o)	&	3	&	322.231	&	1.81E-04	&	3.77E+05	&	4.20E+05	&	-3.23	&	-3.18	&	0.221	\\
	&		&		&		&	39115	&	(o)	&	1	&	412.092	&	1.83E-06	&	3.81E+03	&	4.25E+03	&	-5.01	&	-4.97	&	0.218	\\
	&		&		&		&	39346	&	(o)	&	2	&	416.045	&	3.25E-07	&	6.78E+02	&	7.56E+02	&	-5.76	&	-5.71	&	0.068	\\
	&		&		&		&	55715	&	(o)	&	1	&	1305.250	&	5.75E-06	&	1.20E+04	&	1.34E+04	&	-3.51	&	-3.47	&	0.207	\\
63529	&	(e)	&	4	&	2.19$^e$	&	27602	&	(o)	&	3	&	278.267	&	7.54E-03	&	2.79E+07	&	3.10E+07	&	-1.49	&	-1.44	&	0.119	\\
	&		&		&		&	27841	&	(o)	&	4	&	280.130	&	3.33E-01	&	1.23E+09	&	1.37E+09	&	0.16	&	0.21	&	0.704	\\
	&		&		&		&	28161	&	(o)	&	3	&	282.663	&	6.60E-01	&	2.44E+09	&	2.71E+09	&	0.47	&	0.51	&	0.697	\\
	&		&		&		&	32350	&	(o)	&	3	&	320.641	&	3.38E-05	&	1.25E+05	&	1.39E+05	&	-3.71	&	-3.67	&	0.014	\\
64367	&	(e)	&	2	&	2.25$^e$	&	26081	&	(o)	&	2	&	261.119	&	5.63E-01	&	1.03E+09	&	1.25E+09	&	0.03	&	0.11	&	0.410	\\
	&		&		&		&	27444	&	(o)	&	2	&	270.754	&	1.32E-02	&	2.41E+07	&	2.93E+07	&	-1.58	&	-1.49	&	0.444	\\
	&		&		&		&	27602	&	(o)	&	3	&	271.923	&	5.47E-04	&	1.00E+06	&	1.21E+06	&	-2.96	&	-2.87	&	0.510	\\
	&		&		&		&	27918	&	(o)	&	1	&	274.276	&	3.89E-04	&	7.11E+05	&	8.64E+05	&	-3.10	&	-3.01	&	0.023	\\
	&		&		&		&	28021	&	(o)	&	2	&	275.057	&	4.30E-04	&	7.87E+05	&	9.56E+05	&	-3.05	&	-2.96	&	0.021	\\
	&		&		&		&	28161	&	(o)	&	3	&	276.120	&	8.53E-03	&	1.56E+07	&	1.90E+07	&	-1.75	&	-1.66	&	0.258	\\
	&		&		&		&	29742	&	(o)	&	1	&	288.728	&	3.03E-02	&	5.55E+07	&	6.74E+07	&	-1.16	&	-1.07	&	0.503	\\
	&		&		&		&	29824	&	(o)	&	2	&	289.412	&	1.85E-02	&	3.38E+07	&	4.11E+07	&	-1.37	&	-1.29	&	0.305	\\
	&		&		&		&	30816	&	(o)	&	1	&	297.967	&	3.20E-01	&	5.85E+08	&	7.11E+08	&	-0.11	&	-0.02	&	0.567	\\
	&		&		&		&	32350	&	(o)	&	3	&	312.247	&	4.32E-02	&	7.90E+07	&	9.60E+07	&	-0.94	&	-0.85	&	0.207	\\
	&		&		&		&	39115	&	(o)	&	1	&	395.902	&	3.40E-05	&	6.22E+04	&	7.56E+04	&	-3.84	&	-3.75	&	0.017	\\
	&		&		&		&	39346	&	(o)	&	2	&	399.549	&	9.57E-05	&	1.75E+05	&	2.13E+05	&	-3.38	&	-3.29	&	0.021	\\
	&		&		&		&	55715	&	(o)	&	1	&	1155.577	&	1.90E-03	&	3.48E+06	&	4.23E+06	&	-1.16	&	-1.07	&	0.207	\\
64706	&	(e)	&	2	&	2.51$^e$	&	26081	&	(o)	&	2	&	258.825	&	2.19E-02	&	3.43E+07	&	4.36E+07	&	-1.46	&	-1.36	&	0.199	\\
	&		&		&		&	27444	&	(o)	&	2	&	268.289	&	8.23E-04	&	1.29E+06	&	1.64E+06	&	-2.86	&	-2.75	&	0.238	\\
	&		&		&		&	27602	&	(o)	&	3	&	269.437	&	6.38E-04	&	9.99E+05	&	1.27E+06	&	-2.97	&	-2.86	&	0.208	\\
	&		&		&		&	27918	&	(o)	&	1	&	271.746	&	8.55E-04	&	1.34E+06	&	1.70E+06	&	-2.83	&	-2.72	&	0.055	\\
	&		&		&		&	28021	&	(o)	&	2	&	272.513	&	2.21E-02	&	3.46E+07	&	4.40E+07	&	-1.42	&	-1.31	&	0.154	\\
	&		&		&		&	28161	&	(o)	&	3	&	273.556	&	1.56E-01	&	2.44E+08	&	3.10E+08	&	-0.56	&	-0.46	&	0.252	\\
	&		&		&		&	29742	&	(o)	&	1	&	285.927	&	1.63E-01	&	2.56E+08	&	3.26E+08	&	-0.50	&	-0.40	&	0.424	\\
	&		&		&		&	29824	&	(o)	&	2	&	286.597	&	5.87E-01	&	9.20E+08	&	1.17E+09	&	0.06	&	0.16	&	0.545	\\
	&		&		&		&	30816	&	(o)	&	1	&	294.984	&	4.05E-02	&	6.35E+07	&	8.07E+07	&	-1.08	&	-0.98	&	0.562	\\
\hline
\end{tabular}
\end{small}
\end{center}
\label{gfsc2}
\end{table}
\end{landscape}

\newpage
\begin{landscape}
\addtocounter{table}{-1}
\begin{table}
\caption{Continued.}
\begin{center}
\begin{small}
\begin{tabular}{rrrrrrrrrrrrrr}
\hline
\multicolumn{3}{c}{Upper level$^a$} & \multicolumn{1}{c}{$\tau_u$ (ns)} & \multicolumn{3}{c}{Lower level$^a$}&
 \multicolumn{1}{c}{$\lambda^b$ (nm)} & \multicolumn{1}{c}{$BF$} & \multicolumn{1}{c}{$gA$ (s$^{-1}$)} & \multicolumn{1}{c}{$gA_{\text{resc}}$ (s$^{-1}$)} & \multicolumn{1}{c}{$\log(gf)$} & \multicolumn{1}{c}{$\log(gf)_{\text{resc}}$} & \multicolumn{1}{c}{CF$^c$} \\
\hline
	&		&		&		&	32350	&	(o)	&	3	&	308.973	&	3.45E-03	&	5.40E+06	&	6.87E+06	&	-2.11	&	-2.01	&	0.222	\\
	&		&		&		&	39115	&	(o)	&	1	&	390.654	&	7.53E-04	&	1.18E+06	&	1.50E+06	&	-2.57	&	-2.46	&	0.021	\\
	&		&		&		&	39346	&	(o)	&	2	&	394.204	&	2.45E-03	&	3.84E+06	&	4.88E+06	&	-2.05	&	-1.94	&	0.024	\\
	&		&		&		&	55715	&	(o)	&	1	&	1111.977	&	1.19E-04	&	1.87E+05	&	2.38E+05	&	-2.46	&	-2.36	&	0.198	\\
74433	&	(e)	&	2	&	3.80$^e$	&	26081	&	(o)	&	2	&	206.751	&	7.13E-02	&	5.98E+07	&	9.38E+07	&	-1.42	&	-1.22	&	0.047	\\
	&		&		&		&	27444	&	(o)	&	2	&	212.746	&	1.00E-03	&	8.40E+05	&	1.32E+06	&	-3.25	&	-3.05	&	0.054	\\
	&		&		&		&	27602	&	(o)	&	3	&	213.467	&	5.58E-05	&	4.68E+04	&	7.34E+04	&	-4.51	&	-4.30	&	0.083	\\
	&		&		&		&	27918	&	(o)	&	1	&	214.914	&	8.51E-07	&	7.14E+02	&	1.12E+03	&	-6.32	&	-6.11	&	0.000	\\
	&		&		&		&	28021	&	(o)	&	2	&	215.394	&	2.11E-04	&	1.77E+05	&	2.78E+05	&	-3.92	&	-3.71	&	0.024	\\
	&		&		&		&	28161	&	(o)	&	3	&	216.045	&	1.07E-02	&	8.94E+06	&	1.40E+07	&	-2.21	&	-2.01	&	0.455	\\
	&		&		&		&	29742	&	(o)	&	1	&	223.689	&	5.46E-04	&	4.58E+05	&	7.19E+05	&	-3.47	&	-3.27	&	0.005	\\
	&		&		&		&	29824	&	(o)	&	2	&	224.099	&	2.39E-05	&	2.00E+04	&	3.14E+04	&	-4.83	&	-4.63	&	0.001	\\
	&		&		&		&	30816	&	(o)	&	1	&	229.195	&	4.05E-03	&	3.40E+06	&	5.34E+06	&	-2.58	&	-2.38	&	0.002	\\
	&		&		&		&	32350	&	(o)	&	3	&	237.551	&	8.23E-01	&	6.90E+08	&	1.08E+09	&	-0.24	&	-0.04	&	0.251	\\
	&		&		&		&	39115	&	(o)	&	1	&	283.056	&	2.83E-03	&	2.37E+06	&	3.72E+06	&	-2.56	&	-2.35	&	0.244	\\
	&		&		&		&	39346	&	(o)	&	2	&	284.916	&	1.05E-02	&	8.83E+06	&	1.39E+07	&	-1.98	&	-1.77	&	0.660	\\
	&		&		&		&	55715	&	(o)	&	1	&	534.098	&	4.13E-04	&	3.46E+05	&	5.43E+05	&	-2.85	&	-2.63	&	0.000	\\
	&		&		&		&	66048	&	(o)	&	2	&	1192.292	&	3.94E-02	&	3.30E+07	&	5.18E+07	&	-0.20	&	0.04	&	0.693	\\
	&		&		&		&	66390	&	(o)	&	1	&	1242.891	&	4.01E-05	&	3.36E+04	&	5.27E+04	&	-3.17	&	-2.91	&	0.227	\\
	&		&		&		&	66493	&	(o)	&	2	&	1259.000	&	6.15E-05	&	5.16E+04	&	8.10E+04	&	-2.97	&	-2.72	&	0.112	\\
	&		&		&		&	66460	&	(o)	&	2	&	1253.786	&	1.25E-03	&	1.05E+06	&	1.65E+06	&	-1.66	&	-1.41	&	0.403	\\
	&		&		&		&	66564	&	(o)	&	3	&	1270.370	&	2.48E-04	&	2.08E+05	&	3.26E+05	&	-2.36	&	-2.10	&	0.374	\\
	&		&		&		&	66584	&	(o)	&	3	&	1273.628	&	8.54E-05	&	7.16E+04	&	1.12E+05	&	-2.82	&	-2.56	&	0.390	\\
	&		&		&		&	67298	&	(o)	&	1	&	1401.037	&	5.56E-05	&	4.66E+04	&	7.31E+04	&	-2.93	&	-2.67	&	0.085	\\
	&		&		&		&	67396	&	(o)	&	2	&	1420.650	&	7.89E-07	&	6.62E+02	&	1.04E+03	&	-4.76	&	-4.50	&	0.003	\\
	&		&		&		&	67744	&	(o)	&	3	&	1494.454	&	2.93E-02	&	2.46E+07	&	3.86E+07	&	-0.15	&	0.11	&	0.695	\\
	&		&		&		&	68498	&	(o)	&	1	&	1684.392	&	5.06E-03	&	4.24E+06	&	6.65E+06	&	-0.82	&	-0.55	&	0.165	\\
76361	&	(e)	&	1	&	1.14	$^e$&	26081	&	(o)	&	2	&	198.888	&	2.24E-04	&	5.75E+05	&	5.91E+05	&	-3.47	&	-3.46	&	0.126	\\
	&		&		&		&	27444	&	(o)	&	2	&	204.362	&	3.68E-04	&	9.44E+05	&	9.70E+05	&	-3.23	&	-3.22	&	0.387	\\
	&		&		&		&	27918	&	(o)	&	1	&	206.362	&	7.57E-02	&	1.94E+08	&	1.99E+08	&	-0.91	&	-0.90	&	0.507	\\
	&		&		&		&	28021	&	(o)	&	2	&	206.804	&	2.30E-01	&	5.88E+08	&	6.04E+08	&	-0.42	&	-0.41	&	0.527	\\
	&		&		&		&	29736	&	(o)	&	0	&	214.412	&	9.25E-05	&	2.37E+05	&	2.43E+05	&	-3.79	&	-3.77	&	0.000	\\
	&		&		&		&	29742	&	(o)	&	1	&	214.439	&	2.71E-04	&	6.93E+05	&	7.12E+05	&	-3.32	&	-3.31	&	0.002	\\
	&		&		&		&	29824	&	(o)	&	2	&	214.816	&	5.31E-05	&	1.36E+05	&	1.40E+05	&	-4.03	&	-4.01	&	0.000	\\
	&		&		&		&	30816	&	(o)	&	1	&	219.494	&	9.17E-05	&	2.35E+05	&	2.41E+05	&	-3.77	&	-3.76	&	0.015	\\
	&		&		&		&	39002	&	(o)	&	0	&	267.597	&	2.33E-01	&	5.96E+08	&	6.12E+08	&	-0.19	&	-0.18	&	0.708	\\
	&		&		&		&	39115	&	(o)	&	1	&	268.407	&	1.73E-01	&	4.44E+08	&	4.56E+08	&	-0.32	&	-0.31	&	0.705	\\
	&		&		&		&	39346	&	(o)	&	2	&	270.079	&	2.84E-01	&	7.28E+08	&	7.48E+08	&	-0.10	&	-0.09	&	0.709	\\
	&		&		&		&	55715	&	(o)	&	1	&	484.233	&	1.12E-06	&	2.88E+03	&	2.96E+03	&	-5.00	&	-4.98	&	0.080	\\
	&		&		&		&	66048	&	(o)	&	2	&	969.440	&	1.10E-05	&	2.83E+04	&	2.91E+04	&	-3.40	&	-3.39	&	0.091	\\
	&		&		&		&	66390	&	(o)	&	1	&	1002.628	&	2.87E-04	&	7.35E+05	&	7.55E+05	&	-1.96	&	-1.94	&	0.402	\\
	&		&		&		&	66493	&	(o)	&	2	&	1013.085	&	4.41E-04	&	1.13E+06	&	1.16E+06	&	-1.76	&	-1.75	&	0.319	\\
	&		&		&		&	66460	&	(o)	&	2	&	1009.706	&	1.72E-04	&	4.40E+05	&	4.52E+05	&	-2.17	&	-2.16	&	0.307	\\
	&		&		&		&	67237	&	(o)	&	0	&	1095.698	&	9.72E-04	&	2.49E+06	&	2.56E+06	&	-1.35	&	-1.34	&	0.575	\\
\hline
\end{tabular}
\end{small}
\end{center}
\label{gfsc2}
\end{table}
\end{landscape}

\newpage
\begin{landscape}
\addtocounter{table}{-1}
\begin{table}
\caption{Continued.}
\begin{center}
\begin{small}
\begin{tabular}{rrrrrrrrrrrrrr}
\hline
\multicolumn{3}{c}{Upper level$^a$} & \multicolumn{1}{c}{$\tau_u$ (ns)} & \multicolumn{3}{c}{Lower level$^a$}&
 \multicolumn{1}{c}{$\lambda^b$ (nm)} & \multicolumn{1}{c}{$BF$} & \multicolumn{1}{c}{$gA$ (s$^{-1}$)} & \multicolumn{1}{c}{$gA_{\text{resc}}$ (s$^{-1}$)} & \multicolumn{1}{c}{$\log(gf)$} & \multicolumn{1}{c}{$\log(gf)_{\text{resc}}$} & \multicolumn{1}{c}{CF$^c$} \\
\hline
	&		&		&		&	67298	&	(o)	&	1	&	1103.071	&	6.87E-04	&	1.76E+06	&	1.81E+06	&	-1.49	&	-1.48	&	0.528	\\
	&		&		&		&	67396	&	(o)	&	2	&	1115.192	&	9.76E-04	&	2.50E+06	&	2.57E+06	&	-1.33	&	-1.32	&	0.472	\\
	&		&		&		&	68498	&	(o)	&	1	&	1271.473	&	3.72E-07	&	9.52E+02	&	9.78E+02	&	-4.64	&	-4.63	&	0.029	\\
	&		&		&		&	75308	&	(o)	&	2	&	9497.064	&	1.05E-07	&	2.69E+02	&	2.76E+02	&	-3.44	&	-3.43	&	0.567	\\
	&		&		&		&	75591	&	(o)	&	2	&	12984.147	&	3.27E-07	&	8.37E+02	&	8.60E+02	&	-2.68	&	-2.66	&	0.581	\\
	&		&		&		&	75651	&	(o)	&	1	&	14082.652	&	2.29E-07	&	5.87E+02	&	6.03E+02	&	-2.76	&	-2.75	&	0.295	\\
	&		&		&		&	75681	&	(o)	&	2	&	14699.495	&	9.76E-07	&	2.50E+03	&	2.57E+03	&	-2.10	&	-2.08	&	0.578	\\
	&		&		&		&	75913	&	(o)	&	2	&	22304.392	&	1.26E-08	&	3.22E+01	&	3.31E+01	&	-3.64	&	-3.61	&	0.035	\\
	&		&		&		&	75952	&	(o)	&	1	&	24447.995	&	4.84E-08	&	1.24E+02	&	1.27E+02	&	-2.97	&	-2.94	&	0.375	\\
	&		&		&		&	75994	&	(o)	&	0	&	27287.372	&	2.90E-08	&	7.44E+01	&	7.64E+01	&	-3.11	&	-3.07	&	0.363	\\
	&		&		&		&	76073	&	(o)	&	1	&	34770.709	&	1.85E-10	&	4.75E-01	&	4.88E-01	&	-5.04	&	-5.05	&	0.031	\\
76589	&	(e)	&	2	&	1.09$^e$	&	26081	&	(o)	&	2	&	197.989	&	1.79E-04	&	7.57E+05	&	8.23E+05	&	-3.35	&	-3.32	&	0.040	\\
	&		&		&		&	27444	&	(o)	&	2	&	203.412	&	7.58E-05	&	3.20E+05	&	3.48E+05	&	-3.70	&	-3.67	&	0.170	\\
	&		&		&		&	27602	&	(o)	&	3	&	204.071	&	1.10E-03	&	4.64E+06	&	5.04E+06	&	-2.54	&	-2.50	&	0.490	\\
	&		&		&		&	27918	&	(o)	&	1	&	205.393	&	3.08E-03	&	1.30E+07	&	1.41E+07	&	-2.08	&	-2.05	&	0.390	\\
	&		&		&		&	28021	&	(o)	&	2	&	205.831	&	4.50E-02	&	1.90E+08	&	2.07E+08	&	-0.92	&	-0.88	&	0.454	\\
	&		&		&		&	28161	&	(o)	&	3	&	206.426	&	2.51E-01	&	1.06E+09	&	1.15E+09	&	-0.17	&	-0.13	&	0.529	\\
	&		&		&		&	29742	&	(o)	&	1	&	213.393	&	1.33E-05	&	5.60E+04	&	6.09E+04	&	-4.42	&	-4.38	&	0.000	\\
	&		&		&		&	29824	&	(o)	&	2	&	213.766	&	4.74E-04	&	2.00E+06	&	2.17E+06	&	-2.86	&	-2.83	&	0.001	\\
	&		&		&		&	30816	&	(o)	&	1	&	218.398	&	1.25E-04	&	5.27E+05	&	5.73E+05	&	-3.42	&	-3.39	&	0.012	\\
	&		&		&		&	32350	&	(o)	&	3	&	225.973	&	1.25E-03	&	5.26E+06	&	5.72E+06	&	-2.39	&	-2.36	&	0.328	\\
	&		&		&		&	39115	&	(o)	&	1	&	266.771	&	1.77E-01	&	7.46E+08	&	8.11E+08	&	-0.10	&	-0.06	&	0.705	\\
	&		&		&		&	39346	&	(o)	&	2	&	268.422	&	5.17E-01	&	2.18E+09	&	2.37E+09	&	0.37	&	0.41	&	0.709	\\
	&		&		&		&	55715	&	(o)	&	1	&	478.932	&	2.35E-05	&	9.91E+04	&	1.08E+05	&	-3.47	&	-3.43	&	0.019	\\
	&		&		&		&	66048	&	(o)	&	2	&	948.425	&	2.61E-05	&	1.10E+05	&	1.20E+05	&	-2.82	&	-2.79	&	0.065	\\
	&		&		&		&	66390	&	(o)	&	1	&	980.166	&	1.23E-05	&	5.20E+04	&	5.65E+04	&	-3.13	&	-3.09	&	0.219	\\
	&		&		&		&	66493	&	(o)	&	2	&	990.157	&	1.52E-04	&	6.43E+05	&	6.99E+05	&	-2.03	&	-1.99	&	0.337	\\
	&		&		&		&	66460	&	(o)	&	2	&	986.929	&	6.09E-05	&	2.57E+05	&	2.79E+05	&	-2.42	&	-2.39	&	0.243	\\
	&		&		&		&	66564	&	(o)	&	3	&	997.176	&	6.97E-04	&	2.94E+06	&	3.20E+06	&	-1.36	&	-1.32	&	0.372	\\
	&		&		&		&	66584	&	(o)	&	3	&	999.182	&	1.61E-04	&	6.78E+05	&	7.37E+05	&	-1.99	&	-1.96	&	0.215	\\
	&		&		&		&	67298	&	(o)	&	1	&	1075.944	&	8.96E-04	&	3.78E+06	&	4.11E+06	&	-1.18	&	-1.15	&	0.593	\\
	&		&		&		&	67396	&	(o)	&	2	&	1087.473	&	2.06E-03	&	8.71E+06	&	9.47E+06	&	-0.81	&	-0.77	&	0.517	\\
	&		&		&		&	67744	&	(o)	&	3	&	1130.199	&	9.83E-05	&	4.15E+05	&	4.51E+05	&	-2.10	&	-2.06	&	0.549	\\
	&		&		&		&	68498	&	(o)	&	1	&	1235.566	&	1.70E-05	&	7.17E+04	&	7.79E+04	&	-2.78	&	-2.75	&	0.138	\\
	&		&		&		&	75308	&	(o)	&	2	&	7803.238	&	4.76E-08	&	2.01E+02	&	2.18E+02	&	-3.74	&	-3.70	&	0.536	\\
	&		&		&		&	75309	&	(o)	&	3	&	7806.711	&	2.44E-07	&	1.03E+03	&	1.12E+03	&	-3.03	&	-2.99	&	0.670	\\
	&		&		&		&	75373	&	(o)	&	3	&	8220.091	&	2.89E-07	&	1.22E+03	&	1.33E+03	&	-2.91	&	-2.87	&	0.502	\\
	&		&		&		&	75552	&	(o)	&	3	&	9642.061	&	1.18E-07	&	5.00E+02	&	5.43E+02	&	-3.16	&	-3.12	&	0.304	\\
	&		&		&		&	75591	&	(o)	&	2	&	10012.694	&	8.81E-09	&	3.72E+01	&	4.04E+01	&	-4.26	&	-4.22	&	0.012	\\
	&		&		&		&	75651	&	(o)	&	1	&	10653.532	&	2.30E-09	&	9.70E+00	&	1.05E+01	&	-4.79	&	-4.75	&	0.011	\\
	&		&		&		&	75681	&	(o)	&	2	&	11002.823	&	1.28E-07	&	5.40E+02	&	5.87E+02	&	-3.02	&	-2.97	&	0.092	\\
	&		&		&		&	75716	&	(o)	&	3	&	11444.422	&	2.21E-06	&	9.33E+03	&	1.01E+04	&	-1.75	&	-1.70	&	0.611	\\
	&		&		&		&	75913	&	(o)	&	2	&	14773.133	&	5.78E-07	&	2.44E+03	&	2.65E+03	&	-2.11	&	-2.06	&	0.445	\\
	&		&		&		&	75952	&	(o)	&	1	&	15683.966	&	1.11E-07	&	4.69E+02	&	5.10E+02	&	-2.78	&	-2.73	&	0.354	\\
\hline
\end{tabular}
\end{small}
\end{center}
\label{gfsc2}
\end{table}
\end{landscape}

\newpage
\begin{landscape}
\addtocounter{table}{-1}
\begin{table}
\caption{Continued.}
\begin{center}
\begin{small}
\begin{tabular}{rrrrrrrrrrrrrr}
\hline
\multicolumn{3}{c}{Upper level$^a$} & \multicolumn{1}{c}{$\tau_u$ (ns)} & \multicolumn{3}{c}{Lower level$^a$}&
 \multicolumn{1}{c}{$\lambda^b$ (nm)} & \multicolumn{1}{c}{$BF$} & \multicolumn{1}{c}{$gA$ (s$^{-1}$)} & \multicolumn{1}{c}{$gA_{\text{resc}}$ (s$^{-1}$)} & \multicolumn{1}{c}{$\log(gf)$} & \multicolumn{1}{c}{$\log(gf)_{\text{resc}}$} & \multicolumn{1}{c}{CF$^c$} \\
\hline
	&		&		&		&	76073	&	(o)	&	1	&	19373.811	&	7.70E-09	&	3.25E+01	&	3.53E+01	&	-3.73	&	-3.70	&	0.278	\\
77387	&	(e)	&	3	&	3.73	$^e$&	26081	&	(o)	&	2	&	194.910	&	2.66E-05	&	3.37E+04	&	5.00E+04	&	-4.71	&	-4.55	&	0.015	\\
	&		&		&		&	27444	&	(o)	&	2	&	200.162	&	8.78E-04	&	1.11E+06	&	1.65E+06	&	-3.17	&	-3.00	&	0.300	\\
	&		&		&		&	27602	&	(o)	&	3	&	200.800	&	2.37E-02	&	3.00E+07	&	4.45E+07	&	-1.74	&	-1.57	&	0.312	\\
	&		&		&		&	27841	&	(o)	&	4	&	201.768	&	2.02E-01	&	2.56E+08	&	3.80E+08	&	-0.80	&	-0.63	&	0.327	\\
	&		&		&		&	28021	&	(o)	&	2	&	202.504	&	1.99E-02	&	2.52E+07	&	3.74E+07	&	-1.81	&	-1.64	&	0.293	\\
	&		&		&		&	28161	&	(o)	&	3	&	203.079	&	1.29E-01	&	1.63E+08	&	2.42E+08	&	-0.99	&	-0.82	&	0.269	\\
	&		&		&		&	29824	&	(o)	&	2	&	210.180	&	1.76E-01	&	2.23E+08	&	3.31E+08	&	-0.83	&	-0.66	&	0.374	\\
	&		&		&		&	32350	&	(o)	&	3	&	221.970	&	4.87E-05	&	6.16E+04	&	9.14E+04	&	-4.34	&	-4.17	&	0.126	\\
	&		&		&		&	39346	&	(o)	&	2	&	262.791	&	1.16E-01	&	1.47E+08	&	2.18E+08	&	-0.81	&	-0.65	&	0.249	\\
	&		&		&		&	66048	&	(o)	&	2	&	881.687	&	2.62E-04	&	3.32E+05	&	4.93E+05	&	-2.39	&	-2.24	&	0.123	\\
	&		&		&		&	66493	&	(o)	&	2	&	917.642	&	8.86E-03	&	1.12E+07	&	1.66E+07	&	-0.83	&	-0.68	&	0.509	\\
	&		&		&		&	66460	&	(o)	&	2	&	914.869	&	6.78E-03	&	8.58E+06	&	1.27E+07	&	-0.95	&	-0.80	&	0.781	\\
	&		&		&		&	66564	&	(o)	&	3	&	923.667	&	6.02E-02	&	7.62E+07	&	1.13E+08	&	0.01	&	0.16	&	0.379	\\
	&		&		&		&	66584	&	(o)	&	3	&	925.388	&	7.61E-02	&	9.63E+07	&	1.43E+08	&	0.11	&	0.26	&	0.792	\\
	&		&		&		&	66719	&	(o)	&	4	&	937.110	&	1.33E-01	&	1.68E+08	&	2.49E+08	&	0.36	&	0.52	&	0.748	\\
	&		&		&		&	67396	&	(o)	&	2	&	1000.629	&	4.64E-02	&	5.87E+07	&	8.71E+07	&	-0.04	&	0.12	&	0.575	\\
	&		&		&		&	67744	&	(o)	&	3	&	1036.689	&	4.70E-05	&	5.94E+04	&	8.81E+04	&	-3.00	&	-2.85	&	0.115	\\
	&		&		&		&	75221	&	(o)	&	4	&	4616.186	&	1.04E-06	&	1.31E+03	&	1.94E+03	&	-3.28	&	-3.21	&	0.152	\\
	&		&		&		&	75308	&	(o)	&	2	&	4808.601	&	2.30E-08	&	2.91E+01	&	4.32E+01	&	-4.90	&	-4.82	&	0.068	\\
	&		&		&		&	75309	&	(o)	&	3	&	4809.920	&	4.80E-08	&	6.07E+01	&	9.01E+01	&	-4.58	&	-4.50	&	0.011	\\
	&		&		&		&	75373	&	(o)	&	3	&	4963.717	&	4.93E-07	&	6.24E+02	&	9.26E+02	&	-3.54	&	-3.47	&	0.150	\\
	&		&		&		&	75390	&	(o)	&	4	&	5006.948	&	1.42E-06	&	1.80E+03	&	2.67E+03	&	-3.07	&	-3.00	&	0.104	\\
	&		&		&		&	75470	&	(o)	&	4	&	5215.252	&	1.24E-06	&	1.57E+03	&	2.33E+03	&	-3.09	&	-3.02	&	0.112	\\
	&		&		&		&	75552	&	(o)	&	3	&	5448.967	&	1.03E-06	&	1.30E+03	&	1.93E+03	&	-3.13	&	-3.07	&	0.538	\\
	&		&		&		&	75561	&	(o)	&	4	&	5474.449	&	1.18E-08	&	1.49E+01	&	2.21E+01	&	-5.05	&	-5.00	&	0.107	\\
	&		&		&		&	75591	&	(o)	&	2	&	5565.388	&	5.53E-07	&	7.00E+02	&	1.04E+03	&	-3.37	&	-3.32	&	0.306	\\
	&		&		&		&	75681	&	(o)	&	2	&	5858.418	&	1.88E-06	&	2.38E+03	&	3.53E+03	&	-2.79	&	-2.74	&	0.571	\\
	&		&		&		&	75716	&	(o)	&	3	&	5981.306	&	1.01E-05	&	1.28E+04	&	1.90E+04	&	-2.04	&	-1.99	&	0.848	\\
	&		&		&		&	75913	&	(o)	&	2	&	6779.698	&	4.40E-07	&	5.56E+02	&	8.25E+02	&	-3.28	&	-3.24	&	0.072	\\
\hline
\end{tabular}
\end{small}
\end{center}
\begin{small}
$^a$ Each level is designated by its value in cm$^{-1}$, its parity ((e) and (o) stand
for even and odd respectively) and its total quantum number, $J$.\\
$^b$ Calculated from the energy levels compiled by NIST \citep{kra15}.
$\lambda > 200~{\rm nm }$ are given in air.\\ 
$^c$ Cancellation factor as defined by \citet{cow81}. A value less than 5\%
indicates a strong cancellation effect on the line strength and the transition
probability could be underestimated.\\
$^d$ TR-LIF measurements by \citet{mar88}.\\
$^e$ TR-LIF measurements. This work.
\end{small}
\label{gfsc2}
\end{table}
\end{landscape}

\newpage
\begin{landscape}
\begin{table}
\caption{\ion{Sc}{ii} lines used in the determination of the solar abundance (${\rm log}~\epsilon$) of scandium.}
\begin{center}
\begin{small}
\begin{tabular}{cccrrrrrrr}
\hline
\multicolumn{3}{c}{Transition} &
\multicolumn{1}{c}{$\lambda^a$ (nm)} & \multicolumn{1}{c}{$\log(gf)_{\text{L\&D}}^b$} &
\multicolumn{1}{c}{$\log(gf)_{\text{resc}}^c$} & \multicolumn{1}{c}{$\Delta \log(gf)^d$} &
\multicolumn{1}{c}{${\rm log}~\epsilon^e$} &
\multicolumn{1}{c}{${\rm log}~\epsilon_{\text{cor}}^f$} & \multicolumn{1}{c}{Weight$^e$}\\
\hline
${\rm 3d^2~^3F_4}$	&	$-$	& ${\rm 3d4p~^3F^o_3}$		& 442.067&	-2.273	&	-1.950	&	0.323	&
	3.099	&	2.776 &  2 	\\
${\rm 3d^2~^3F_3}$	&	$-$	& ${\rm 3d4p~^3F^o_2}$		& 443.135&	-1.969	&	-1.780	&	0.189	&
	3.155	&	2.966 &  1 	\\
${\rm 3d^2~^3P_2}$	&	$-$	& ${\rm 3d4p~^1P^o_1}$		& 535.720&	-2.111	&	-2.050	&	0.061	&
	3.131	&	3.070 &  2 	\\
${\rm 3d^2~^3P_1}$	&	$-$	& ${\rm 3d4p~^3P^o_2}$		& 564.100&	-1.131	&	-1.000	&	0.131	&
	3.226	&	3.095 &  1 	\\
${\rm 3d^2~^3P_0}$	&	$-$	& ${\rm 3d4p~^3P^o_1}$		& 565.836&	-1.208	&	-1.150	&	0.058	&
	3.211	&	3.153 &  1 	\\
${\rm 3d^2~^3P_1}$	&	$-$	& ${\rm 3d4p~^3P^o_1}$		& 566.715&	-1.309	&	-1.220	&	0.089	&
	3.235	&	3.146 &  1 	\\
${\rm 3d^2~^3P_1}$	&	$-$	& ${\rm 3d4p~^3P^o_0}$		& 566.904&	-1.200	&	-1.100	&	0.100	&
	3.246	&	3.146 &  1 	\\
${\rm 3d^2~^3P_2}$	&	$-$	& ${\rm 3d4p~^3P^o_1}$		& 568.420&	-1.074	&	-1.030	&	0.044	&
	3.154	&	3.110 &  2 	\\
${\rm 3d^2~^1D_2}$	&	$-$	& ${\rm 3d4p~^1D^o_2}$		& 660.460&	-1.309	&	-1.570$^g$ 	&	-0.261	&
	3.204	&	3.465 &  1 	\\
\hline
\end{tabular}
\end{small}
\end{center}
\begin{small}
$^a$ Calculated from the energy levels compiled by NIST \citep{kra15}.\\ 
$^b$  \citet{sco15}, deduced from the $A$-values of \citet{law89}.\\
$^c$ This work. Rescaled value. \\
$^d$ $\Delta \log(gf) = \log(gf)_{\text{L\&D}} - \log(gf)_{\text{resc}}$. \\
$^e$ \citet{sco15}.\\
$^f$ This work. Corrected abundance, ${\rm log}~\epsilon_{\text{cor}} = {\rm log}~\epsilon - \Delta \log(gf)$.\\
$^g$ Affected by strong cancellation effects ($CF<0.05$).
\end{small}
\label{abun}
\end{table}
\end{landscape}







\bsp	
\label{lastpage}
\end{document}